\documentclass[]{aa}

\usepackage{amsmath}
\usepackage{amssymb}
\usepackage{url}
\usepackage{graphicx}
\usepackage{times}
\usepackage[utf8]{inputenc}
\usepackage{xspace}
\usepackage{color}
\usepackage{multirow}
\usepackage{multicol}
\usepackage{booktabs}
\usepackage{float}
\usepackage{mathtools, cuted}

\bibpunct{(}{)}{;}{a}{}{,}

\usepackage{etoolbox}
\makeatletter
\patchcmd\@combinedblfloats{\box\@outputbox}{\unvbox\@outputbox}{}{%
  \errmessage{\noexpand\@combinedblfloats could not be patched}%
}%
\makeatother

\def\apjl{ApJL}
\def\apjs{ApJSS}

\def\aap{A\&A}

\def\nothing{}

\def\tj{\theta_\mathrm{j}}

\def\ba{\beta_\mathrm{a}}
\def\Ga{\Gamma_\mathrm{a}}
\def\rhoa{\rho_\mathrm{a}}
\def\rhoj{\rho_\mathrm{j}}
\def\bj{\beta_\mathrm{j}}
\def\Gj{\Gamma_\mathrm{j}}

\def\bh{\beta_\mathrm{h}}
\def\Gh{\Gamma_\mathrm{h}}

\newcommand{\simpropto}{\mathrel{\vcenter{
  \offinterlineskip\halign{\hfil$##$\cr
    \propto\cr\noalign{\kern2pt}\sim\cr\noalign{\kern-2pt}}}}}

\title{Gamma-ray burst jet propagation, development of angular structure, and the luminosity function}
\titlerunning{GRB jet propagation and structure development}
\author{O.~S.~Salafia\inst{\ref{oab.me},\ref{infn.mib}}\thanks{E-mail: om.salafia@inaf.it} \and C.~Barbieri\inst{\ref{unimib},\ref{infn.mib}} \and S.~Ascenzi\inst{\ref{gssi},\ref{oab.me}} \and M.~Toffano\inst{\ref{oab.me},\ref{unins}}}

\institute{INAF -- Osservatorio Astronomico di Brera, via E. Bianchi 46, I-23807 Merate, Italy\label{oab.me} \and Università degli Studi di Milano-Bicocca, Dip. di Fisica ``G. Occhialini'', Piazza della Scienza 3, I-20126 Milano, Italy\label{unimib} \and INFN -- Sezione di Milano-Bicocca, Piazza della Scienza 3, I-20126 Milano, Italy\label{infn.mib} \and Gran Sasso Science Institute, viale F. Crispi 7, I-67100 L'Aquila (AQ), Italy\label{gssi} \and Università degli Studi dell'Insubria, Via Valleggio 11, 22100, Como, Italy\label{unins}}

\authorrunning{O.~S.~Salafia, C.~Barbieri, S.~Ascenzi and M.~Toffano}

\date{Received xxx / Accepted: xxx}

\abstract{
The fate and observable properties of gamma-ray burst jets depend crucially on their interaction with the progenitor material that surrounds the central engine. We present a semi-analytical model of such interaction, which builds upon several previous analytical and numerical works, aimed at predicting the angular distribution of jet and cocoon energy and Lorentz factor after breakout, given the properties of the ambient material and of the jet at launch. Using this model, we construct synthetic populations of structured jets, assuming either a collapsar (for long gamma-ray bursts -- LGRBs) or a binary neutron star merger (for short gamma-ray bursts -- SGRBs) as progenitor. We assume all progenitors to be identical, and we allow little variability in the jet properties at launch: our populations therefore feature a quasi-universal structure. These populations are able to reproduce the main features of the observed LGRB and SGRB luminosity functions, although several uncertainties and caveats remain to be addressed.
}

\keywords{relativistic processes, gamma-ray burst:general, stars:neutron}

\begin{document}

\maketitle

\section{Introduction}

By the time the connection between long gamma-ray bursts (GRBs) and core-collapse supernovae had been finally established (with the association of SN1998bw to GRB980425 -- \citealt{Galama1998,Patat2001} -- and that of SN2003dh to GRB030329 -- \citealt{Hjorth2003,Stanek2003,Matheson2003}) the community started to realise that long GRB (LGRB) jets must confront themselves with the stellar envelope before being able to expand in the interstellar space up to transparency. This early phase of interaction with the ambient material can leave imprints on the jet properties, whose modeling can be used to constrain the properties of the progenitor \citep{Matzner2003}. Numerical simulations \citep[e.g.][]{Aloy2000,Zhang2003,Morsony2007,Xie2019} and (semi-)analytical models \citep[e.g.][]{Ramirez-Ruiz2002,Matzner2003,Morsony2007,Bromberg2011} have been used to investigate the dynamics of this interaction. In the following years, evidence accumulated in favour of the idea that short GRBs (SGRBs) had a different progenitor \citep[see e.g.][whose deep upper limits could exclude the presence of a supernova associated to GRB050709]{Fox2005}, the most favoured option being that of a compact object merger involving at least one neutron star \citep{Narayan1992}. The first investigation of the interaction of a SGRB jet with a post-neutron star merger environment was presented by \citet{Aloy2005}, whose numerical simulations found the effect of the interaction with the environment to be significant on the jet propagation and on the determination of its structure. Subsequent, increasingly refined simulations \citep[e.g.][]{Nagakura2014,Just2016,Gottlieb2018,Xie2018,Duffell2018,Geng2019} confirmed these results, showing that in some cases the merger ejecta could be even dense enough as to choke the jet.

All these simulations (regardless of the progenitor type) also showed that the distribution of energy per unit solid angle and the average Lorentz factor of the jet after breakout are strong functions of the angle from the jet axis, i.e.~realistic jets are structured (as opposed to the uniform -- or ``top-hat'' -- jet widely assumed in GRB modelling) and the structure is determined by both the properties of the jet at launch and its interaction with the progenitor ambient medium. \citet{Lipunov2001} and \citet{Rossi2002} were the first to realise that the presence of jet structure could lead to a radically different appearance of GRB jets when seen under different viewing angles, and that this could be one of the main effects to shape the GRB luminosity function. The LGRB luminosity function was later shown to be consistent with a single quasi-universal jet structure with a uniform core and a steep decrease of energy density away from the jet axis \citep{Pescalli2015,Salafia2015}. \citet{Kumar2003}, \citet{Granot2003} and \citep{Rossi2004} were the first to provide a detailed modelling of the afterglow emission from a structured jet, showing that the most prominent differences, with respect to the uniform jet model, are for off-axis observers. Structured jets have been invoked to explain some particular LGRBs (such as GRB 030329,  \citealt{Berger2003}) and more recently in the context of SGRBs as electromagnetic counterparts of gravitational wave events \citep[e.g.][]{Lazzati2016,Lamb2017a,Kathirgamaraju2017,Salafia2017b}.

Despite these advancements, the uniform jet model prevailed until very recently for its simplicity and for its success in describing at least a fraction of GRB afterglows. The first compelling evidence of a structured jet, indeed, came only from observations of GRB170817A \citep[e.g.][]{Ghirlanda2019,DAvanzo2018,Lazzati2018,Lyman2018,Mooley2018,Margutti2018,Resmi2018,Troja2018}, the GRB associated to GW170817, the first neutron star merger detected in gravitational waves by LIGO and Virgo \citep{Abbott2017a,Abbott2017b}. This has been the first GRB jet to be conclusively observed off-axis, despite an associated prompt emission was detected. Its afterglow multi-wavelength light curves, the apparently superluminal motion \citep{Mooley2018} and small projected size \citep{Ghirlanda2019} of its radio image, and the late time steep decay \citep{Lamb2019,Mooley2018b} could only be explained invoking a structured jet seen under a $15$--$25^\circ$ viewing angle. Intriguingly, both the prompt and afterglow emission of a jet with the same structure, located at the median distance and within the median interstellar medium density of the previously known SGRB population, if seen on-axis, would have fallen right in the middle of the known SGRB population \citep{Salafia2019}, hinting at a quasi-universal SGRB jet structure.

In this work we present a semi-analytical model of the interaction of the jet with the ambient medium, which represents an attempt at identifying and modelling the main aspects of this complex physical process. We then use this model to construct synthetic populations of LGRB and SGRB jets, under the assumption that the properties of these jets at launch and those of their progenitors vary little within the population. This leads naturally to a quasi-universal structure. The comparison of the luminosity distributions of these synthetic populations with the observation-based reconstructed luminosity functions of GRBs shows a general agreement. This provides support to the idea that GRB jets share a quasi-universal structure, even though several uncertainties and caveats remain to be addressed.

\section{Model of the jet propagation through the ambient medium}\label{sec:jet_propagation}

\subsection{Initial remarks}

\nothing{In this and the following section, we describe our model for the jet propagation through, and breakout from, the ambient medium. The model is purely hydrodynamic, i.e.~we do not include a description of the effects of the magnetic field. Since the main candidate mechanism for jet launching \citep{Blandford1977} requires a strong magnetic field at the base, this may seem a major oversight. The propagation of a Poynting-flux-dominated jet into an ambient medium can be in principle much different from that of a hydrodynamic jet \citep[e.g.][]{Komissarov1999,Lyubarsky2009,Bromberg2014}. On the other hand, as initially argued by \citet{Levinson2005} based on analytical considerations, and later demonstrated by \citet{Bromberg2016} by means of relativistic MHD simulations, when such a Poynting-flux-dominated jet propagates in a dense ambient medium and starts being collimated by the cocoon, it becomes unstable to internal kink modes, which lead to magnetic reconnection at the collimation nozzle. This dissipates the magnetic field down to equipartition, leading to a propagation that resembles that of a purely hydrodynamic, hot jet \citep{Levinson2005,Bromberg2016}. External kink modes can still cause the jet head to oscillate in the transverse direction, effectively increasing the jet working surface and therefore slowing down the propagation \citep{Bromberg2016}, but this effect is progressively reduced during the propagation as long as the ambient material density profile decreases faster than $z^{-2}$ (which should always be the case in GRB progenitors). We thus expect only a minor impact of this simplification on our conclusions, and we defer a more detailed consideration of the inclusion of the magnetic field to a future work.}

\begin{figure}
    \centering
    \includegraphics[width=\columnwidth]{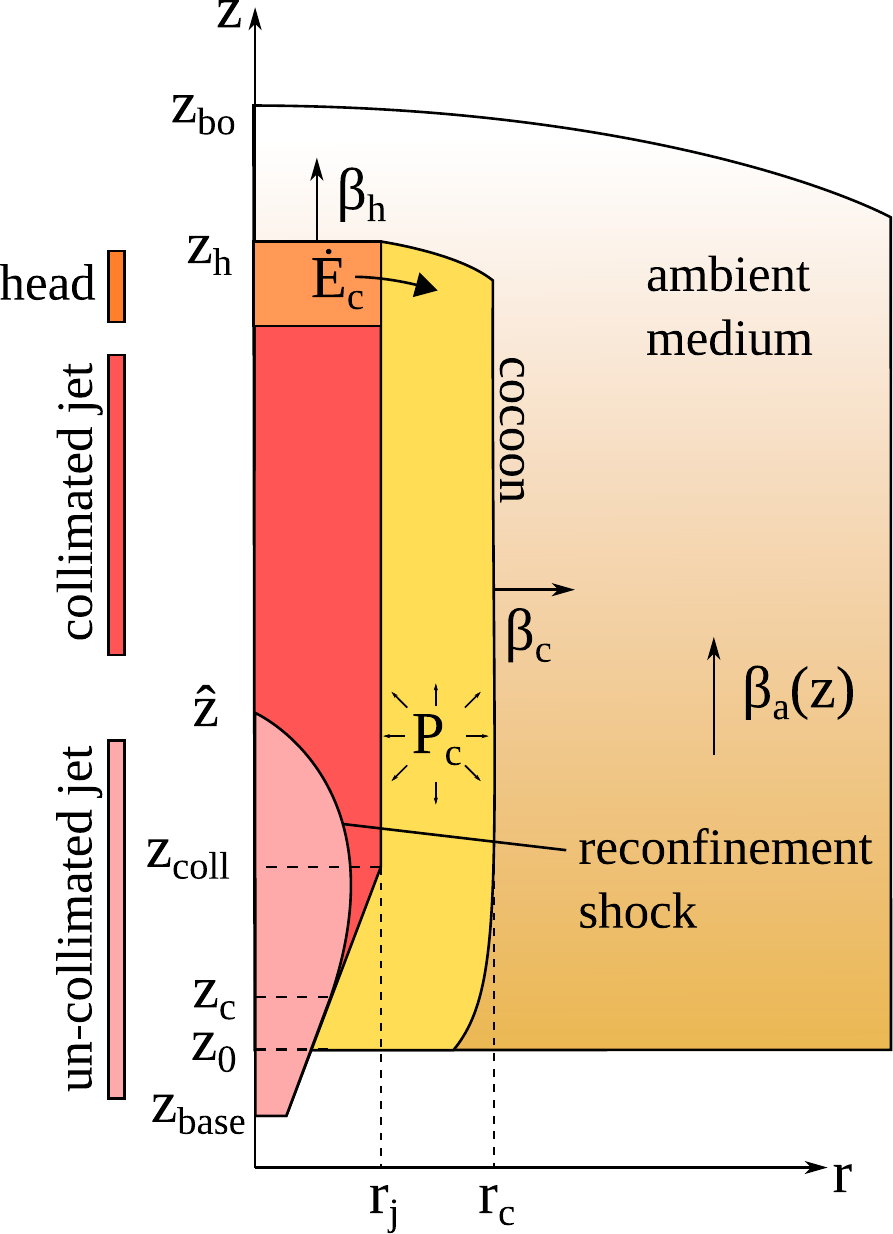}
    \caption{Sketch of the various components of the jet propagation model, with some key quantities annotated. All quantities are defined and described in the text.}
    \label{fig:setting}
\end{figure}

\nothing{Let us therefore} consider a relativistic jet launched by some central engine located at the origin of our coordinate system. The jet is represented by an outflow that initially moves radially within a cone of half-opening angle $\theta_\mathrm{j,0}$, directed towards the $z$ axis (let us employ cylindrical coordinates), with a constant luminosity $L_\mathrm{j}$. The base of the jet is located at a height $z_{base}$. If the central engine is surrounded by some ambient material, the jet will collide with it and form a forward shock (that propagates into the ambient medium) and a reverse shock (where the jet material enters the shocked region). We call ``head'' the region comprised between the forward and reverse shock. Figure~\ref{fig:setting} shows a sketch of the described setting, with key quantities annotated, to be used as reference throughout this section.

\subsection{Jet head advancement}

As shown by e.g.~\cite{Marti1995} and \cite{Matzner2003}, the jet head velocity can be estimated by balancing the ram pressure of the jet and that of the ambient material in the head rest frame. In the case of a moving ambient medium, \cite{Murguia-Berthier2017} showed 
that the resulting head velocity is given by
\begin{equation}
\bh = \frac{\ba + \bj\sqrt{\tilde{L}}}{1+\sqrt{\tilde{L}}}
\label{eq:beta_head}
\end{equation}
where 
\begin{equation}
    \tilde{L} = \frac{\rhoj'h_\mathrm{j}'\Gj^2}{\rhoa'\Ga^2} = \frac{L_\mathrm{j}}{\pi \theta_\mathrm{j}^2 z^2 \rhoa \Gamma_\mathrm{a} c^3}.
    \label{eq:Ltilde}
\end{equation}
Here the subscripts $\mathrm{a}$ and $\mathrm{j}$ indicate respectively the ambient and jet material, primed quantities are measured in the comoving frame of the respective fluid, $\beta$ indicates the velocity in units of the speed of light $c$, $\Gamma=(1-\beta^2)^{-1/2}$ is the Lorentz factor, $\rho$ is the rest mass density, $h$ is the specific dimensionless enthalpy, and $\tj$ is the jet head half-opening angle. 

Numerical relativistic hydrodynamics simulations generally show that the jet head proceeds slower than predicted by Eq.~\ref{eq:beta_head} in presence of jet collimation by the ambient medium. As detailed in \cite{Harrison2018}, the agreement with simulations in the case of a static ambient medium can be restored by replacing $\tilde L$ in Eq.~\ref{eq:beta_head} with the effective value $\tilde L_\mathrm{eff}=A^2(\tilde L,\theta_\mathrm{j,0},\Omega)\tilde L$, where
\begin{equation}
    A(\tilde L,\theta_\mathrm{j,0},\Omega) = \left\lbrace\begin{array}{ll}
        0.35 & \tilde L \leq 1  \\
        0.35 \tilde L^{\frac{0.46}{\log_{10}\tilde L_\mathrm{col}}} & 1 < \tilde L \leq \tilde L_\mathrm{col} \\
        1 & \tilde L > \tilde L_\mathrm{col}\\
    \end{array}\right.
    \label{eq:harrison_corr}
\end{equation}
where $\tilde L$ is computed using Eq.~\ref{eq:Ltilde} evaluated at the jet head, and $\tilde L_\mathrm{col}=\theta_\mathrm{j,0}^{-4/3}(16 \Omega/3)^{2/3}$. Here $\Omega$ is a functional of the ambient medium density profile whose exact form is described in \cite{Harrison2018}.
In what follows, we employ this correction, assuming it to be valid also in the case of a moving ambient medium.

\subsubsection{The cocoon}

As the jet head advances, the swept ambient material is cast aside, forming an over-pressured region surrounding the jet, which is usually called ``cocoon''. We assume the energy flow from the head to the cocoon $\dot E_\mathrm{c}$ to be a fraction $\eta$ of the jet energy flow through the reverse shock, that is $\eta L_\mathrm{j}(\beta_\mathrm{j}-\beta_\mathrm{h})$. We estimate $\eta(t)$ as the fraction of the head volume that is in causal contact with the cocoon at a given time $t$. Let us assume the sound speed in the head to be $c_\mathrm{s}=c/\sqrt{3}$. The time-scale associated to the head advancement, as measured in the head comoving frame, is $t_\mathrm{h}\sim z_\mathrm{h}/\bh\Gh c$, so the region of the head in causal contact with the cocoon extends to a distance $l\sim z_\mathrm{h} /\sqrt{3}\bh\Gh$ inside the head. Assuming the head to be a cylinder of radius $r_\mathrm{j}=\tj z_\mathrm{h}$, the fraction of its volume being in causal contact with the cocoon at a given time is then given by
\begin{equation}
    \eta = \left\lbrace\begin{array}{lr}
        \frac{2}{\mu}-\frac{1}{\mu^2} & \mu>1  \\
        1 & \mu\leq 1\\ 
    \end{array}\right.
    \label{eq:eta}
\end{equation}
where $\mu=\sqrt{3}\tj\Gh\bh$.
This is similar to the prescription given in \cite{Bromberg2011}, with the advantage that it is a continuous function of the jet head velocity.

We assume the cocoon to be at rest in a frame that moves upwards at a speed $\bar{\beta}_\mathrm{a}$ which is the rest-mass averaged speed of the ambient medium from $z_0$ (the height of its base) to $z_\mathrm{h}$, namely
\begin{equation}
    \bar{\ba} = \frac{\int_{z_0}^{z_\mathrm{h}}\rhoa(z) \ba(z)dz}{\int_{z_0}^{z_\mathrm{h}}\rhoa(z)dz}
\end{equation}
and we define $\bar\Gamma_\mathrm{a}=(1-\bar\ba^2)^{-1/2}$.
The energy of the cocoon is  given by
\begin{equation}
    E_\mathrm{c}(t)=\int_{0}^t \eta(t)L_\mathrm{j}(\bj-\bh(t))\,dt
    \label{eq:cocoon_energy}
\end{equation}
where $t$ is measured in the central engine rest frame and the jet luminosity is evaluated at the retarded time $t-z_\mathrm{h}/\bj c$. The term in parentheses accounts for the relative velocity between the jet and the head. 


\subsubsection{Cocoon pressure}

We assume the cocoon to be cylindrical, extending from the base of the ambient material located at a height $z_0$ (which could be above the jet base if a cavity is present around the central engine, and moves with the ambient material if the wind stops being injected from the central region) and extending up to the jet head at $z_\mathrm{h}$ (where the head bow shock is present). Let us indicate the cocoon cylindrical radius as $r_\mathrm{c}$. We assume the cocoon to be radiation-dominated and neglect pressure gradients, so that its pressure is $P_\mathrm{c}=E_\mathrm{c}/3\pi r_\mathrm{c}^2 (z_\mathrm{h}-z_0)\bar\Gamma_\mathrm{a}$. The cocoon expands sideways at a speed $\beta_c$ which is obtained by balancing the cocoon pressure and the ram pressure from the external medium. To keep with our assumption of cylindrical cocoon shape, we average out this speed vertically, namely we write $\beta_\mathrm{c}=\left(1+\bar\rho_\mathrm{a}c^2/P_\mathrm{c}\right)^{-1/2}$, where $\bar\rho_\mathrm{a}=(z_\mathrm{h}-z_0)^{-1}\int_{z_0}^{z_\mathrm{h}}\rhoa(z,t)dz$ is the altitude-averaged ambient density.

\subsubsection{Reconfinement shock}\label{sec:reconfinement}
The jet is surrounded by the cocoon, which exerts an approximately uniform lateral pressure $P_\mathrm{c}$ inwards on the jet. If this pressure is higher than thermal pressure in the jet $P_\mathrm{j}\sim L z_\mathrm{base}^2/4\pi \theta_\mathrm{j,0}^2 z^4 c$ \citep[][assuming an adiabatic jet]{Matzner2003}, a reconfinement shock will form \citep{Komissarov1997,Morsony2007}. The shape of the shock is given by the balance between the component of the jet ram pressure normal to the shock and the cocoon pressure. In the frame of the ambient material, within small angles from the jet axis, the balance equation is given by
\begin{equation}
    \rhoj'h_\mathrm{j}'\Gj^2\Ga^2(\bj-\ba)^2\Ga^{-2}\left(\frac{r_s}{z}-\frac{dr_s}{dz}\right)^2+P_\mathrm{j}=P_\mathrm{c}
\end{equation}
where $r_s$ is the cylindrical radius of the reconfinement shock, and $z$ is measured in the central engine frame (the $\Ga^{-2}$ factor above accounts for its transformation to the ambient medium frame). Assuming $P_\mathrm{c}\gg P_\mathrm{j}$ we neglect the latter and write the solution of the above ordinary differential equation as
\begin{equation}
    r_s(z)=\theta_\mathrm{j,0}(1+A z_c)z - \theta_\mathrm{j,0}A z^2
\end{equation}
where $A=\sqrt{\pi c \bj P_\mathrm{c}/ L_\mathrm{j}(\bj-\ba)^2}$, and $z_c$ is the height at which the jet thermal pressure equals the cocoon pressure \citep{Bromberg2011}. The reconfinement shock thus converges to the jet axis at a height $\hat{z}=A^{-1}+z_\mathrm{c}$. Following \cite{Bromberg2011}, owing to the parabolic shape of the reconfinement shock, we assume the jet cross section radius $r_\mathrm{j}$ to be constant above $(\hat{z}+z_c)/2$, being given by $r_\mathrm{j}=\theta_\mathrm{j,0}(\hat{z}+z_c)/2$. For simplicity, any variation in the conditions (especially the cocoon pressure) near $z_\mathrm{c}$ is assumed to affect instantaneously the reconfinement shock and thus the jet head opening angle.

\section{Jet breakout and development of jet structure}\label{sec:Jet_structure}

\subsection{Breakout condition}\label{sec:breakout_condition}
\nothing{Most realistic progenitor density profiles feature a smooth decrease of the mass density beyond some radius, therefore we need an unambiguous way of defining the breakout radius. Since the jet head forward shock is radiation-dominated, it remains confined to the progenitor interior as long as the Thomson optical depth $\tau(z_\mathrm{h})=\int_{z_\mathrm{h}}^\infty (\sigma_\mathrm{T}/m_\mathrm{p})\rho_\mathrm{a}(z)\,dz$ from the head height $z_\mathrm{h}$ outwards is larger than unity}\footnote{\nothing{For relativistic shocks, the confinement remains effective out to a lower optical depth $\tau\sim 0.01\, \Gamma_\mathrm{s}^3$, where $\Gamma_\mathrm{s}$ is the shock Lorentz factor \citep{Nakar2012}. The shock energy release at breakout, though, still happens mostly at the radius for which $\tau\sim 1$. Given the steep density decrease in the outermost radii of our progenitor models, changing the prescription would anyway translate to a negligible difference in the breakout radius.}}\nothing{(here $\sigma_\mathrm{T}$ and $m_\mathrm{p}$ indicate the Thomson cross section and the proton mass respectively). We therefore take $\tau=1$ as our breakout condition.}

\subsection{Post-breakout evolution}
As soon as the jet head \nothing{breaks out} of the ambient medium (i.e.~the progenitor star in the long GRB scenario, or the outer edge of the compact binary merger ejecta in the short GRB scenario), at a lab-frame time $t_\mathrm{bo}$ and a height $z_\mathrm{bo}$, both the jet and the cocoon are free to expand in the interstellar medium (ISM). As the cocoon material starts to flow out of the new open channel, the shock at its sides stalls and its pressure drops, making the jet collimation at its base soon ineffective \citep{Lazzati2005}. Since the jet in our model is effectively collimated at a height $z_\mathrm{coll}=(\hat{z}+z_c)/2$, the information on the cocoon pressure loss starts affecting the jet opening angle only after a delayed time $t_\mathrm{delay}=(z_\mathrm{bo}-z_\mathrm{coll})/c_\mathrm{s}$, where $c_\mathrm{s}=c/\sqrt{3}$ is the sound speed\footnote{In the treatment described in \cite{Lazzati2005}, this time delay is not considered.}. After this time, we assume the cocoon pressure at $z_\mathrm{coll}$ to drop exponentially as
\begin{equation}
 P_\mathrm{c}(t) = P_\mathrm{c,bo}\exp\left(-\frac{c_\mathrm{s}(t-t_\mathrm{bo}-t_\mathrm{delay})}{z_\mathrm{bo}}\right)
\end{equation}
due to the cocoon material flowing at sound speed out of the open channel. 
As detailed in \citet{Lazzati2005}, the pressure drop causes the jet half-opening angle to increase exponentially
\begin{equation}
 \theta_\mathrm{j}(t) = \theta_\mathrm{j,bo}\exp\left(\alpha\frac{c_\mathrm{s}(t-t_\mathrm{bo}-t_\mathrm{delay})}{z_\mathrm{bo}}\right)
 \label{eq:theta_j(t)}
\end{equation}
until it eventually reaches the base jet half-opening angle $\theta_\mathrm{j,0}$, or until the jet injection stops. Here $\theta_\mathrm{j,bo}$ is the half-opening angle of the head at breakout
, and $\alpha$ is a parameter that depends on the details of the transient acceleration phase of the jet after breakout \citep{Lazzati2005}. We set this value to $1/8$ (it was $1/4$ in \citealt{Lazzati2005}) based on a comparison with numerical simulations (\S\ref{sec:comparison_simulations_collimation}). \nothing{Material that breaks out within an angle $\theta_\mathrm{j}(t)<\Gamma_\mathrm{j}^{-1}$ will subsequently expand laterally to $\Gamma_\mathrm{j}^{-1}$ \citep{Lazzati2005}, so we actually take $\theta_\mathrm{j}(t)$ as the maximum between the value computed from Eq.~\ref{eq:theta_j(t)} and $\Gamma_\mathrm{j}^{-1}$.}

This simple modelling neglects the entrainment (and the possible development of instabilities) between the jet and the surrounding cocoon material, which can transfer some of the jet energy outwards. As a simple effective description of this transfer, we assume that the jet energy that flows out of the progenitor (after breakout) during each time interval $dt$ is spread over the latitudinal angle in a Gaussian fashion, with a sigma equal to $\theta_\mathrm{j}(t)$. The final jet structure is then obtained by integrating over $t$, namely
\begin{equation}
 \frac{dE_\mathrm{jet}}{d\Omega}(\theta) = \int_{t_\mathrm{bo}}^{T_\mathrm{jet}+z_\mathrm{bo}/\beta_\mathrm{j}c}L_\mathrm{j}\,\mathcal{G}(\theta_\mathrm{j}(t),\theta)\,dt
 \label{eq:dEjet/dOmega}
\end{equation}
where
\begin{equation}
 \mathcal{G}(\theta_\mathrm{j},\theta) \propto \exp\left[-\frac{1}{2}\left(\frac{\theta}{\theta_\mathrm{j}}\right)^2\right]\cos^2\theta
 \label{eq:angular_spread_function}
\end{equation}
is the assumed angular energy spread function, and $2\pi\int_0^1 \mathcal{G}(\theta_\mathrm{j},\theta)d\cos\theta=1$. The $\cos^2\theta$ term is inserted to ensure that the function goes to zero as $\theta\to\pi/2$. The quantity $T_\mathrm{jet}$ in Eq.~\ref{eq:dEjet/dOmega} represents the jet duration in the central engine frame, so that $T_\mathrm{jet}+z_\mathrm{bo/\beta_\mathrm{j}c}$ is the time when the jet ceases to flow out of the ambient medium. With this description, a jet that propagates uncollimated by the ambient material develops a Gaussian structure $dE/d\Omega \simpropto \exp[-(\theta/\theta_\mathrm{j,0})^2/2]$. Similarly a jet that is collimated, but whose duration is short compared to the timescale $t_\mathrm{\theta} = t_\mathrm{delay} + 8 z_\mathrm{bo}/c_\mathrm{s}$ over which the jet cross section increases, will have $dE/d\Omega \simpropto \exp[-(\theta/\theta_\mathrm{j,bo})^2/2]$. Cases in between will feature a shallower decrease followed by a Gaussian cut-off. For both short and long GRBs, typical values of $t_\mathrm{\theta}$ are longer than the jet duration, thus the typical GRB jet structure is most likely narrow with a steep fall off outside the jet core.
 
\section{Cocoon structure}\label{sec:Cocoon_structure}

After breakout, the shocked ambient material in the cocoon is free to expand under the effect of its own internal pressure. As discussed in the preceding section, initially its material will blow out of the open channel formed after the jet breakout, and then expand (anisotropically) in the surrounding space. Based on the standard theory of radiation-dominated relativistic fireballs \citep{Cavallo1978,Piran1993}, we can estimate its average terminal Lorentz factor as $\bar\Gamma_\mathrm{c}\sim 1+E_\mathrm{c}/M c^2$, where 
\begin{equation}
 M = \int_{0}^{t_\mathrm{bo}}\pi r_\mathrm{j}^2(t)\rho_\mathrm{a}(z_\mathrm{h}(t)) (\beta_\mathrm{h}(t)-\beta_\mathrm{a}(t))c\,dt
 \label{eq:Mcocoon}
\end{equation}
is the mass in the inner cocoon (i.e.~the ambient mass swept and cast aside as the jet head propagates). Initially, it will expand within an angle $\sim r_\mathrm{c,bo}/z_\mathrm{bo}$, where $r_\mathrm{c,bo}$ is the cocoon cylindrical radius at breakout, but its internal pressure will cause it to expand laterally to an angle $\theta_\mathrm{c}\sim \max\left(r_\mathrm{c,bo}/z_\mathrm{bo},\mathrm{arcsin}(\bar\Gamma_\mathrm{c}^{-1})\right)$. 
These simple arguments give us the Lorentz factor and angular scale of the cocoon structure. 

\subsection{Energy angular structure}

Numerical relativistic hydrodynamical simulations \citep{Lazzati2017,Lazzati2019} suggest that the cocoon typically features an approximately exponential distribution of energy per unit solid angle. Inspired by that, we assume the following ansatz energy distribution
\begin{equation}
 \frac{dE_\mathrm{c}}{d\Omega}(\theta) = K(E_\mathrm{c}) \exp\left(-\frac{\theta}{\theta_\mathrm{c}}\right)\cos^2\theta
 \label{eq:dEcocoon/dOmega}
\end{equation}
where $\theta_\mathrm{c}$ is defined in the preceding section, and $K(E_\mathrm{c})$ is chosen so that the total kinetic energy in the cocoon is $E_\mathrm{c}$, namely
\begin{equation}
 K(E_\mathrm{c})=\frac{E_\mathrm{c}}{2\pi \int_0^1\exp\left(-\frac{\theta}{\theta_\mathrm{c}}\right)\cos^2\theta\, d(\cos\theta)}
\end{equation}
which implies the assumption that all internal energy is converted to kinetic energy.

\subsection{Lorentz factor angular structure}\label{sec:Gamma_structure}
We still need to define how the average Lorentz factor varies with the angular distance from the jet axis. We assume the jet material, where not mixed with the cocoon, to reach a terminal Lorentz factor $\Gamma_\mathrm{j}$ (which is a free parameter of the model). Based on the natural expectation that the cocoon material closer to the jet axis will move faster than that at larger angles, we assign to the cocoon material the ansatz Lorentz factor profile (similar to \citealt{Lazzati2019})
\begin{equation}
 \Gamma_\mathrm{c}(\theta)-1 = (\bar\Gamma_\mathrm{c}-1)\exp\left(-\omega\frac{\theta}{\theta_\mathrm{c}}\right)
\end{equation}
where $\bar\Gamma_\mathrm{c}$ and $\theta_\mathrm{c}$ have been defined in the preceding section, and $\omega$ is a free parameter which we set to $\omega=1/3$ based on comparison with numerical simulations (\S\ref{sec:comparison_simulations_structure}) 

\section{Final structure}

After breakout, the jet and cocoon form a single structure (often referred to as a structured jet), which soon reaches homologous (i.e.~ballistic) expansion. While some rearrangement of the internal structure is still possible in this phase, we assume it to be negligible and simply compute the final angular energy distribution as the sum of the jet and cocoon energies, namely
\begin{equation}
    \frac{dE}{d\Omega}(\theta) = \frac{dE_\mathrm{jet}}{d\Omega}(\theta) + \frac{dE_\mathrm{c}}{d\Omega}(\theta)
\end{equation}
For what concerns the Lorentz factor of the structured jet, we compute it as the mass-weighted average of those of the jet and cocoon material, namely
\begin{equation}
 \Gamma(\theta) = \frac{\frac{dE_\mathrm{jet}}{d\Omega}(\theta)+\frac{dE_\mathrm{c}}{d\Omega}(\theta)}{\Gamma_\mathrm{j}^{-1}\frac{dE_\mathrm{jet}}{d\Omega}(\theta)+\Gamma_\mathrm{c}^{-1}(\theta)\frac{dE_\mathrm{c}}{d\Omega}(\theta)}
\end{equation}
\section{The luminosity distribution of GRBs from structured jets}

Equipped with the model presented in the preceding sections, we proceed to constructing a synthetic population of GRB jets, each with its own structure set by the interaction with the ambient medium. We treat long GRBs (LGRBs) and short GRBs (SGRBs) as separate, independent populations. For each, we choose a single, representative progenitor model, into which we inject jets whose properties are distributed within a narrow range. This results in a population of structured jets (those which successfully punch through the ambient medium) and choked jets (those which do not). Based on a simple modelling of the prompt emission, we compute the luminosity distribution of each population, assuming isotropic viewing angles. Finally, we compare the results with the luminosity functions derived from the observations.

\subsection{Prompt emission model}\label{sec:prompt_emission_model}

In order to construct our luminosity distributions, we need to model the jet prompt emission isotropic-equivalent peak luminosity $L_\mathrm{iso}$ that results from a given jet structure, as a function of the viewing angle. We proceed as follows:
\begin{itemize}
 \item given the jet injection duration $T_\mathrm{jet}$, we compute the GRB duration as $T_\mathrm{GRB} = T_\mathrm{jet} - (t_\mathrm{bo}-z_\mathrm{bo}/\beta_\mathrm{j}c)$ and we assume it to be independent of the viewing angle. This assumption is based on the idea that the gamma-ray emission is powered by some form of energy dissipation within the jet (e.g.~internal shocks or magnetic reconnection), and that it consists of a series of short episodes taking place at some typical radius. Under these assumptions, the duration of the GRB is set by the post-breakout jet injection duration, and does not depend strongly on the viewing angle \citep[see e.g.][]{Salafia2016}. As a caveat, we note that this requires the angular timescale $R/\Gamma^2 c$ at the photospheric radius $R$ (at the angles that contribute significantly to the emission for a given viewing angle) to be shorter than the duration of the burst itself, otherwise the angular time spread would dominate;
 \item we compute the prompt emission isotropic equivalent energy $E_\mathrm{iso}(\theta_\mathrm{v})$ following \citet{Salafia2015} and \citet{Salafia2019}, that is, we assume 10 percent of the kinetic energy at each angle to be converted into gamma-ray radiation, and we integrate the emission over the jet accounting for relativistic beaming;
 \item we assume the prompt emission light curve to have a triangular shape, so that $L_\mathrm{iso}(\theta_\mathrm{v})=2 E_\mathrm{iso}(\theta_\mathrm{v})/T_\mathrm{GRB}$. This is a crude approximation, given the diversity of GRB light curves (especially in LGRBs), but we adopt it for its simplicity.
 
\end{itemize}

For the present study, we only apply this model to cases where the jet successfully breaks out of the progenitor. We defer the implementation of more detailed emission models and the inclusion of the possible prompt emission in the case of a choked jet to future work. 

\subsection{Long GRBs}

\subsubsection{Ambient medium density profile}

\begin{figure}
 \includegraphics[width=\columnwidth]{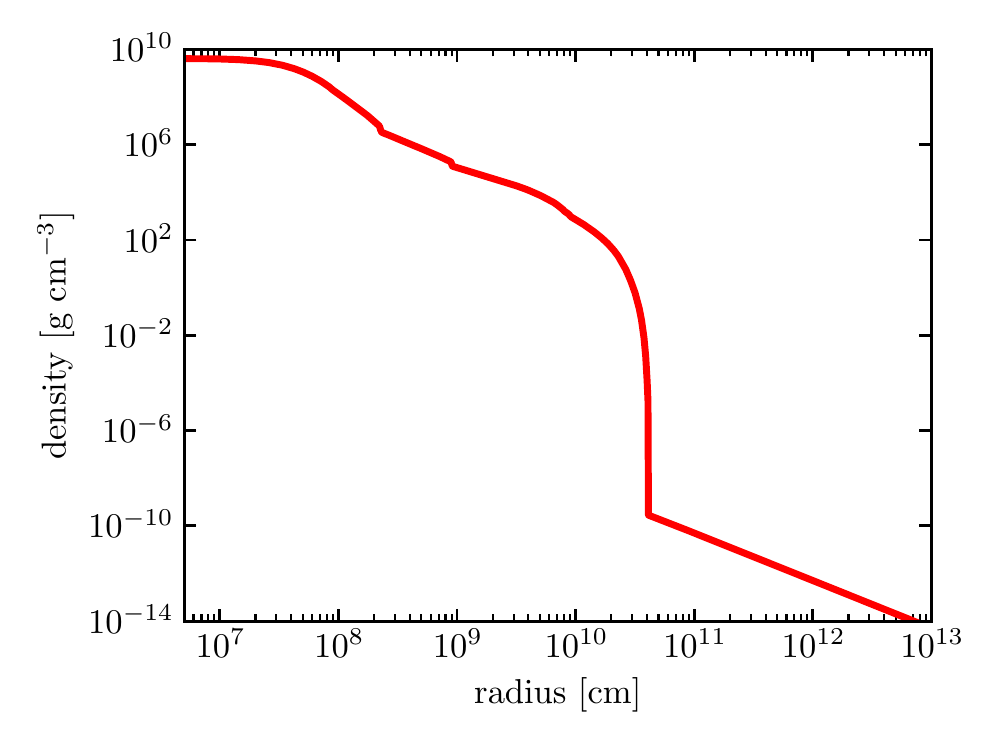}
 \caption{Radial density profile from our representative LGRB progenitor (model 16TI of \citealt{Woosley2006}), which represents a low-metallicity, high-angular momentum, massive star prior to collapse. \nothing{The low-density power-law tail, visible in the lower-right corner of the figure, has been added by ourselves to the original density profile of the model. It represents a stellar wind with typical Wolf-Rayet mass loss rate $\dot M_\mathrm{w}=10^{-5}\mathrm{M_\odot/yr}$ and velocity $v_\mathrm{w}=1000\,\mathrm{km/s}$.}}
 \label{fig:16TI}
\end{figure}

For our representative LGRB progenitor ambient medium, we take the density profile from the stellar model 16TI of \citet{Woosley2006}. This model has been used in several previous studies of LGRB jets \citep[e.g.][]{Morsony2007,LopezCamara2013}. It represents a star with an initial mass of $M=16\,\mathrm{M_\odot}$, low metallicity $Z=10^{-2}\,\mathrm{Z_\odot}$, and a large initial angular momentum $J= 3\times 10^{52}\,\mathrm{erg\;s}$, which is evolved to pre-core collapse using the \texttt{KEPLER} code \citep{Weaver1978}. \nothing{The original simulation data\footnote{Retrieved from \url{https://2sn.org/GRB2/}} features a low-mass ($\Delta M \sim 0.1\,\mathrm{M_\odot}$) extended ($\Delta R\sim 7\times 10^{11}\,\mathrm{cm}$) envelope -- represented by a single point in the adaptive grid -- which is considered \citep{Woosley2006} an artifact due to the treatment of rotation in a one-dimensional simulation. We therefore remove this point from the density profile; we add, for consistency, a stellar wind with typical Wolf-Rayet mass loss rate $\dot M_\mathrm{w}=10^{-5}\mathrm{M_\odot/yr}$ and velocity $v_\mathrm{w}=1000\,\mathrm{km/s}$, even though this has a negligible impact on the modelled jet dynamics. The final density profile is shown in Figure~\ref{fig:16TI}. We note that our results, as detailed below, do not depend strongly on the detailed shape of the progenitor density profile: we obtain similar results employing the progenitor model of \citet{Duffell2015}. A more systematic investigation of the dependence on the progenitor properties will be presented in a future work.}

\subsubsection{Jet properties at launch}\label{sec:LGRB_injection_properties}



Given the large uncertainties on the jet launching mechanism and on its initial colimation, we simply assume all jets to have a fixed half opening angle $\theta_\mathrm{j,0}=0.25\,\mathrm{rad}\approx 14^\circ$ at injection. We also assume their terminal Lorentz factor to be $\Gamma_\mathrm{j}=100$ in all cases (while higher values are actually observed in some cases, this seems to be a typical value at least for LGRBs, see \citealt{Ghirlanda2017}). We extract the jet luminosity $L_\mathrm{j}$ from a log-normal distribution \nothing{centered at $\mu=2\times 10^{49}\,\mathrm{erg/s} \equiv \left <L \right > $ with a dispersion $\sigma=0.65\,\mathrm{dex}$}. Similarly, we extract the duration from a log-normal distribution with $\mu=30\,\mathrm{s} \equiv \left <T \right > $ and $\sigma=0.45\,\mathrm{dex}$. These values are educated guesses, based on the fact that the typical collimation-corrected kinetic energy of LGRBs is around $10^{51}\mathrm{erg}$ \citep{Goldstein2016}, and their average rest-frame duration is just below $30\,\mathrm{s}$ \citep{Salafia2015}.  We choose $z_\mathrm{base}=5\times 10^{6}\,\mathrm{cm}$ as our injection height. We simulate in total 1000 jets, 94\% of which successfully break out of the progenitor star. 

\subsubsection{Resulting LGRB structures}

\begin{figure}
 \includegraphics[width=\columnwidth]{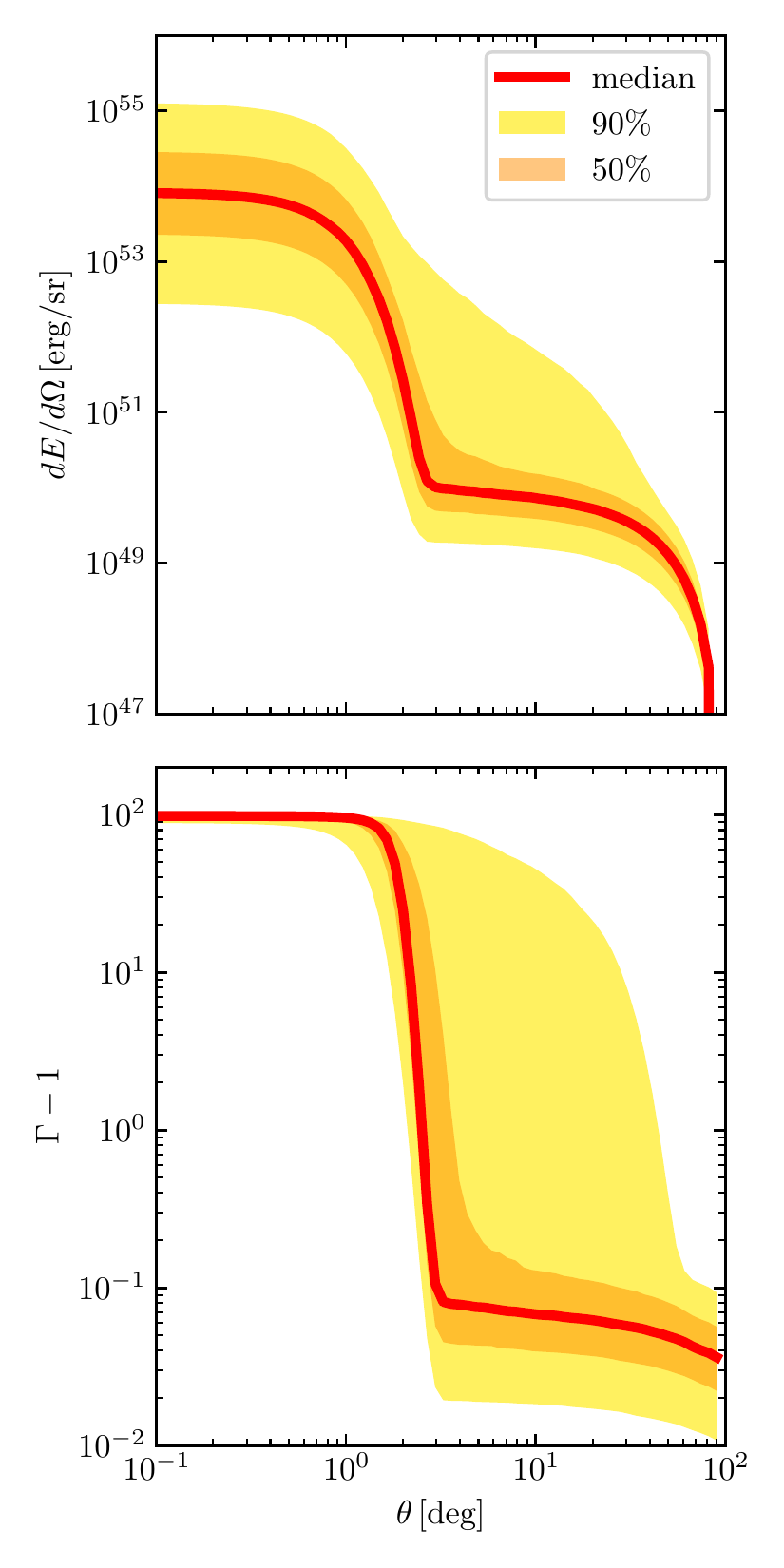}
 \caption{Distribution of jet structures in our LGRB population. In each panel, at each fixed angle $\theta$, the yellow (resp.~orange) filled area encloses 90\% (resp.~50\%) of the structures in the population, while the red line shows the median of the distribution at that angle. The top panel refers to the energy structures (kinetic energy per unit solid angle as a function of the angle $\theta$ from the jet axis), while the bottom panel refers to the velocity structures (average Lorentz factor diminished by one, as a function of $\theta$). }
 \label{fig:struct_distrib_LGRB}
\end{figure}

Figure~\ref{fig:struct_distrib_LGRB} shows the distribution of jet structures for the successful jets in our synthetic LGRB population. The red solid line represents the median value of the kinetic energy per unit solid angle (upper panel) and $\Gamma-1$ (lower panel, respectively) at a given angle $\theta$. The orange and yellow shaded regions contain 50\% and 90\% of the values at each fixed $\theta$, respectively. The figure shows that our LGRB jets are on average very narrow, featuring a core of $\lesssim 1.5\,\mathrm{deg}$, with a steep fall off of the kinetic energy density outside. The cocoon is in general quite energetic, but wide, so that its typical energy density per unit solid angle is lower by around four orders of magnitude with respect to the jet core. The very narrow jet opening angle might seem at odds with those derived from jet breaks in LGRB afterglow light curves \citep{Berger2014}, but we caution that (i) the jet structure (especially if narrow, see e.g.~\citealt{Granot2012}) is expected to evolve after breakout due to jet lateral spread, and it may thus be different at the afterglow stage (see also \citealt{Gill2019b}), and (ii) in the context of structured jet light curves, jet-break-like features are related to the viewing angle rather than to the jet opening angle \citep{Kumar2003,Rossi2004}. Interestingly, the general features we find for the LGRB jet structure -- a narrow opening angle and a faster angular decrease in energy than in Lorentz factor -- are in agreement with the observational constraints derived by \citet{Beniamini2019b}.

\subsubsection{Luminosity distribution}

\begin{figure*}
 \includegraphics[width=\textwidth]{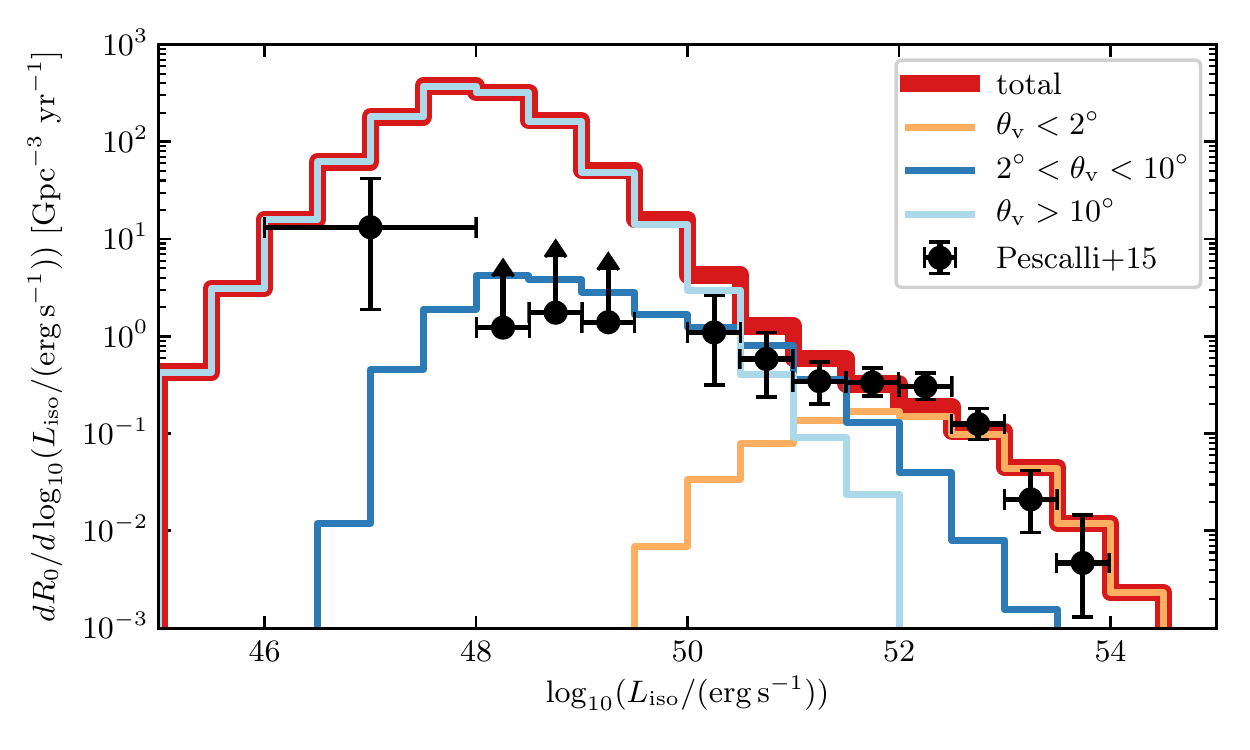}
 \caption{LGRB luminosity distribution of our model population (thick solid red histogram) compared to the observed distribution (black points, from \citealt{Pescalli2015}). Thinner histograms show the luminosity distributions of three sub-classes corresponding to three viewing angle bins, reported in the legend.}
 \label{fig:L_distrib}
\end{figure*}

Figure~\ref{fig:L_distrib} shows the luminosity distribution of our LGRB population, computed as described in the preceding sections. The thick solid red histogram shows the local ($z=0$) rate density per logarithmic luminosity bin, $dR_0/d\log_{10}(L_\mathrm{iso}/(\mathrm{erg\,s^{-1}}))$, constructed from our model population, assuming a \nothing{total local rate} $R_0=600\,\mathrm{yr^{-1}\,Gpc^{-3}}$. The thin, coloured histograms show the luminosity distributions of jets seen within \nothing{2 deg} from the jet axis (yellow), between \nothing{2 and 10 deg} (dark blue) and at viewing angles larger than \nothing{10 deg} (light blue). The black points show the actual luminosity distribution of LGRBs from \citet{Pescalli2015}, who collected and updated binned rate estimates from previous works: the points to the right of $L_\mathrm{iso}=10^{50}\,\mathrm{erg/s}$ come from the reconstructed luminosity function from \citet{Wanderman2010}; the leftmost point is the low-lumnosity GRB rate from \citet{Soderberg2004}, updated as described in \citet{Pescalli2015}; the three remaining points are lower limits derived in \citet{Pescalli2015}. The general features of the model popoulation are in good agreement with the observed points, which is remarkable given the fact that we based our parameter distributions on educated guesses
This shows that the luminosity distribution of LGRBs can be successfully interpreted in a scenario where both jets and progenitors have very similar properties, the main parameter behind the diversity being the viewing angle.

\subsubsection{Dependence on the parameters}

\begin{figure*}[t]
    \centering
    \includegraphics[width=\textwidth]{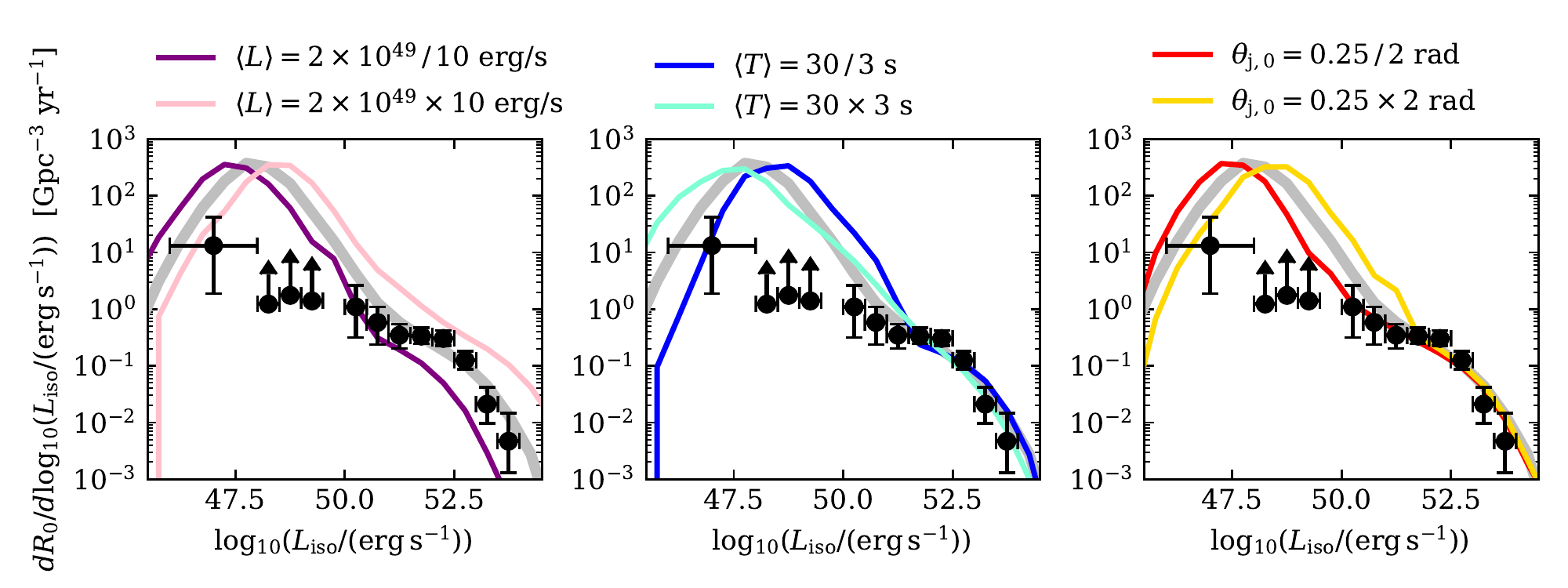}
    \caption{\nothing{Long GRB luminosity function dependence on the parameters of the jet at launch. Each panel shows how the fiducial luminosity function (thick grey solid line) is affected by a change in one of the parameters (coloured thinner lines). See text for an interpretation of the dependencies.}}
    \label{fig:long_param_dependence}
\end{figure*}

\nothing{Figure~\ref{fig:long_param_dependence} gives a grasp of how the luminosity function depends on the jet properties at launch. The grey thick line is the fiducial luminosity function (the same as that shown in Fig.~\ref{fig:L_distrib}). Thinner coloured lines show the variation induced by changing one of the parameters of the distributions of jet properties at launch at a time. To produce each line we construct a synthetic population of 1000 jets whose properties are extracted from distributions with the same parameters as those described in \S\ref{sec:LGRB_injection_properties}, except one, which is displaced as reported in the legend. The resulting luminosity functions (always constructed only from jets that successfully break out) are normalized to the same total rate as the fiducial one. The general trends, and some tentative qualitative explanations, are the following:}
\begin{itemize}
    \item \nothing{increasing the average jet luminosity $\left < L \right >$ by a factor of $10$ (pink line in the left-hand panel of Fig.~\ref{fig:long_param_dependence}) essentially shifts the whole luminosity function to higher luminosities, keeping the shape similar;}
    \item \nothing{lowering $\left < L \right >$ by the same factor (purple line in the left-hand panel of Fig.~\ref{fig:long_param_dependence}), the high-luminosity end is shifted towards lower luminosities, but the low-luminosity end remains essentially unchanged: this is the combined effect of a lower average luminosity after breakout, and of an increased ratio of cocoon to jet energy, which makes the structure shallower (see next point);}
    \item \nothing{lowering the average jet injection duration $\left < T \right >$ (blue line in the central panel of Fig.~\ref{fig:long_param_dependence}) or increasing the jet opening angle at launch (yellow line in the right-hand panel) both make the luminosity function slope below $L_\mathrm{iso}\sim 10^{51.5}\,\mathrm{erg/s}$ steeper. This is due to the larger energy in the cocoon, which makes the structure, broadly speaking, shallower \citep{Pescalli2015}. Also the opposite is true, as shown by the remaining lines.}
\end{itemize}

\nothing{A more detailed account of the dependence of the luminosity function on the parameters and assumptions will be presented in a future work.}

\subsection{Short GRBs}

\subsubsection{Ambient medium density profile}

\begin{figure}
 \includegraphics[width=\columnwidth]{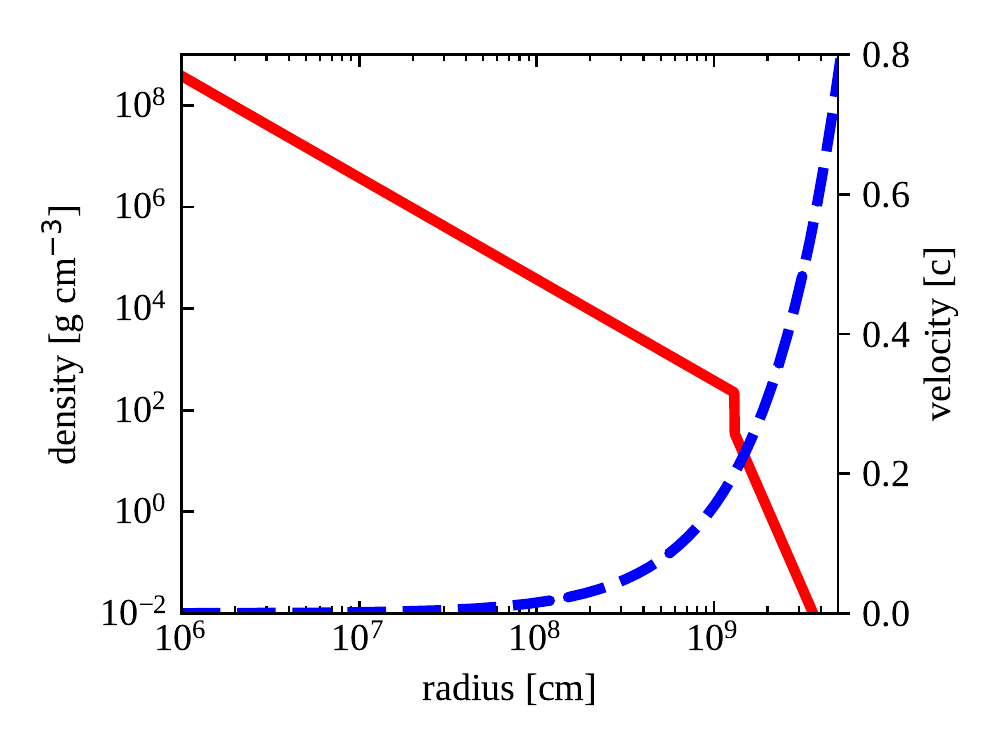}
 \caption{Density and velocity profile of the SGRB ambient medium model. The red line shows a snapshot of the density profile (along the polar axis) of the neutron star merger ejecta model described in \citet{Xie2018} (also used in \citealt{Kasliwal2017} and \citealt{Gottlieb2018}). The blue dashed line shows the corresponding velocity profile (velocity axis on the right). The ejecta are assumed to expand homologously.}
 \label{fig:Xie_ambient}
\end{figure}

Assuming that all (or most) SGRBs are produced by the merger of two neutron stars, and that there is little variation (among different events) in the properties of the various outflows produced during and after the merger, we assume the homologously expanding ejecta cloud described in \citet{Xie2018} to be representative of the SGRB ambient medium. Current SGRB observations seem to actually indicate some degree of diversity in the associated kilonova emission \citep{Gompertz2018}, which would in turn suggest a variety outflow properties, but it is difficult at present to quantify it. We therefore stick with the single progenitor ambient medium model for simplicity. The density profile along the jet axis (solid red line) and velocity profile (dashed blue line) are shown in Figure~\ref{fig:Xie_ambient}.

\subsubsection{Jet properties at injection}\label{sec:SGRB_injection_properties}



As for the LGRB population, we assume all jets to have a fixed half opening angle $\theta_\mathrm{j,0}=0.25\,\mathrm{rad}$ at injection and a terminal Lorentz factor $\Gamma_\mathrm{j}=100$. We extract the jet luminosity $L_\mathrm{j}$
from a log-normal distribution centered at $\mu=3\times 10^{49}\,\mathrm{erg/s}$ with a dispersion $\sigma=0.85\,\mathrm{dex}$; we extract the duration from a log-normal distribution \nothing{with} $\mu=0.3\,\mathrm{s}$ and $\sigma=0.45\,\mathrm{dex}$. For SGRBs, since the ambient medium is not static, we need to define an additional parameter, that is the delay between the neutron star merger (which corresponds to the time when all the ambient material is concentrated at $r=0$) and the start of the jet injection. This represents the time it takes for the post-merger system to develop the necessary conditions to launch a jet (see \citealt{Gill2019} for an interesting discussion of the delay between GW170817 and GRB170817A and its implications for that system). These conditions are very uncertain, as the actual jet launching mechanism is still debated, but the most likely option seems to be energy extraction from a spinning black hole surrounded by an accreting torus of highly magnetized material, through the mechanism first described in \citealt{Blandford1977} \citep[see also][]{Tchekhovskoy2010}. For this to happen, (i) the merger remnant must have collapsed to a black hole, (ii) the magnetic field must have been amplified in the torus (mainly by magneto-rotational instabilities) to reach a significant magnetisation, and (iii) an ordered large scale magnetic field structure with a significant poloidal component must have developed. The time for (ii) is likely very short (few dynamical times), as shown by high-resolution GRMHD simulations (e.g.~\citealt{Kiuchi2018,Kawamura2016}). Process (iii) should take a few Alfvén times \citep[tens of milliseconds, see e.g.][whose simulations resolve the magnetic field amplification, its large scale organization and the consequent jet launching]{Christie2019}. In absence of a prompt collapse to a BH (which would anyway leave little matter outside), therefore, the delay is likely dominated by (i). If the merger remnant is a hypermassive proto-neutron star (which is supported by differential rotation), the collapse to a BH takes tens to hundreds of milliseconds; on the other hand, if the merger remnant is a supra-massive neutron star (which is supported by solid-body rotation), the collapse to a BH takes place only after electromagnetic spin down slows down the remnant enough for it to become unable to support itself against self-gravity: in this case, the collapse could take place after several seconds. It is likely, though, that no accretion disk would be left after such a late collapse \citep{Margalit2015}.  The most likely scenario for a successful SGRB jet to be launched therefore seems that of a short-lived hypermassive proto-neutron star \citep[see also][]{Shibata2006}. Based on these arguments, we fix our time delay at $\Delta t_\mathrm{inj}=0.1\,\mathrm{s}$. Of the total 1000 jets we simulate, 88\% successfully break out of the ambient medium. 

\subsubsection{Distribution of SGRB structures}

\begin{figure}
 \includegraphics[width=\columnwidth]{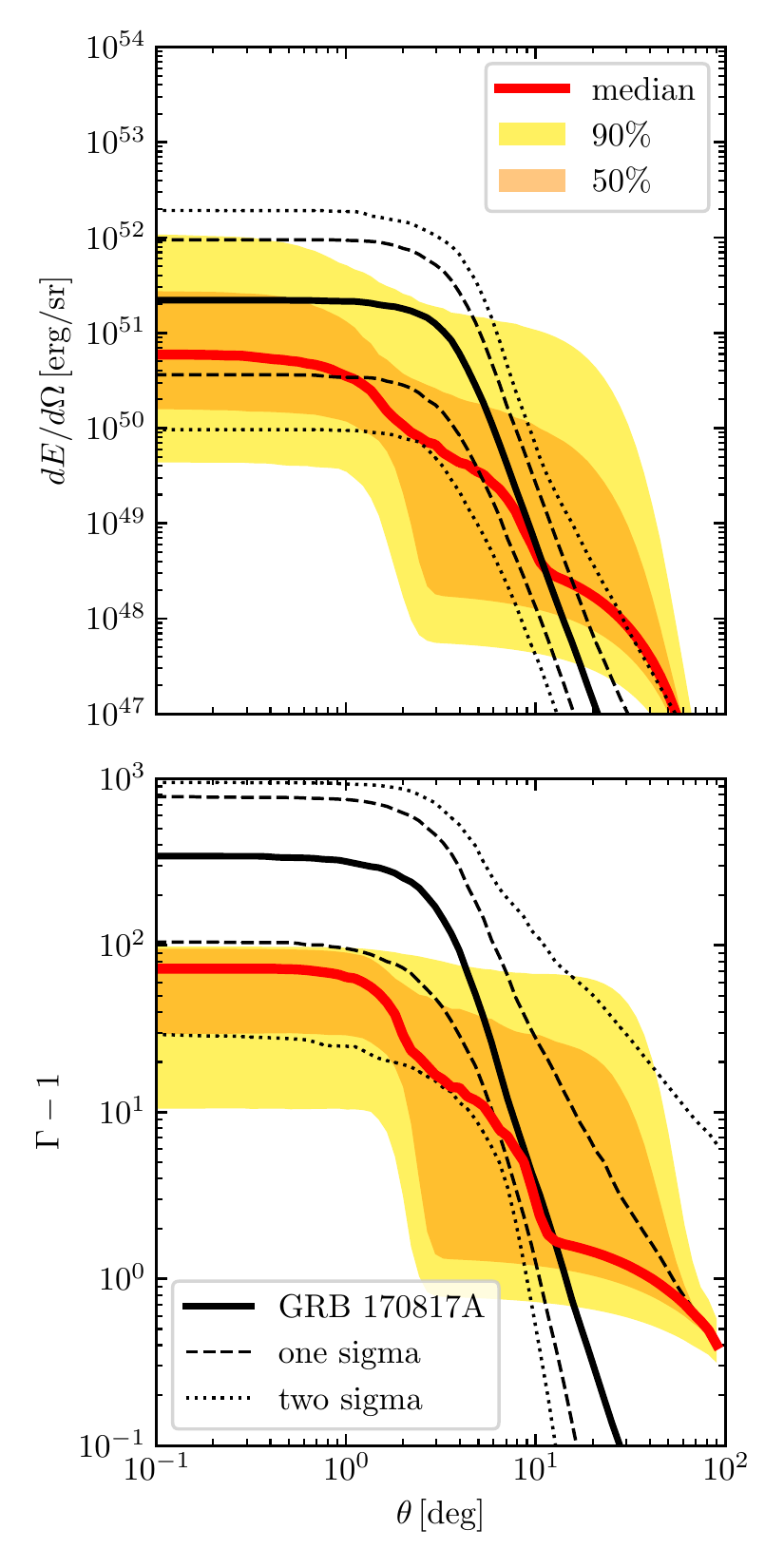}
 \caption{Distribution of structures in our SGRB population. The filled areas and the red lines have the same meaning as in Fig.~\ref{fig:struct_distrib_LGRB}. The black lines show the confidence regions (dotted: two sigma, dashed: one sigma, solid: best fit) for the jet structure of GRB 170817A from \citet{Ghirlanda2019}.}
 \label{fig:struct_distrib_SGRB}
\end{figure}

The jet structure distribution for our synthetic SGRB population features a somewhat larger dispersion with respect to the LGRB population, as shown in Figure~\ref{fig:struct_distrib_SGRB}. It is interesting to compare this distribution with the jet structure of GRB 170817A derived by \citet{Ghirlanda2019} based on multi-wavelength fitting of the afterglow light curves and of the centroid motion observed in VLBI images. The black lines in Fig.~\ref{fig:struct_distrib_SGRB} represent the best-fit (solid), one sigma (dashed) and two sigma (dotted) contours of the derived jet structure. These show that a good fraction of jet structures in our synthetic population are compatible\footnote{The large core Lorentz factor in the structure of GRB 170817A from \citet{Ghirlanda2019} is simply the result of the fact that the self-similar nature of the jet deceleration makes it impossible to distinguish jet core Lorentz factors larger than $\sim \theta_\mathrm{v}^{-1}$ when the jet is observed off-axis, and thus essentially any core Lorentz factor larger than about 30 is compatible with the observations.} with GRB 170817A.

\subsubsection{Luminosity distribution}

\begin{figure*}
 \includegraphics[width=\textwidth]{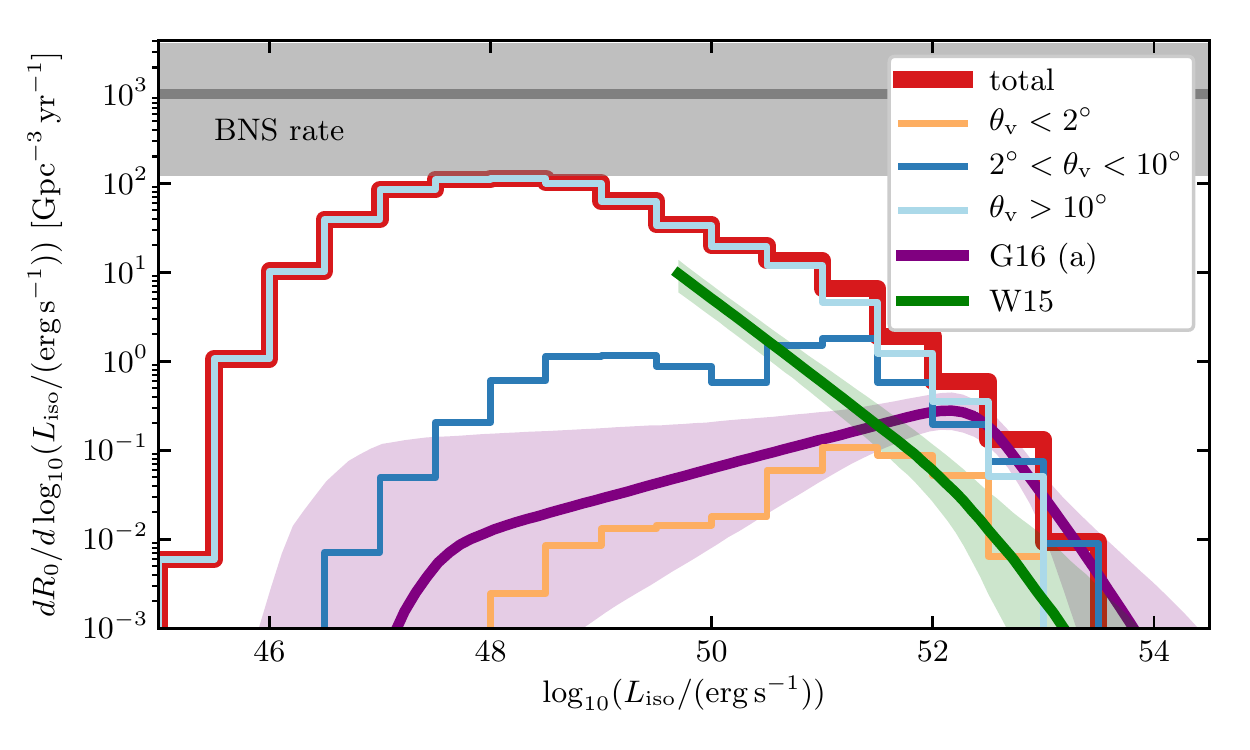}
 \caption{SGRB luminosity distribution of our model population (thick solid red histogram) compared to two luminosity function models based on observations (\citealt{Wanderman2015} -- green solid line and shaded area -- and \citealt{Ghirlanda2016} -- purple solid line and shaded area). Thinner histograms show the luminosity distributions of sub-classes of jets belonging to three viewing angle bins, reported in the legend. The grey line and filled horizontal band show the binary neutron star merger local rate based on gravitational wave observations \citep{TheLIGOScientificCollaboration2018}.}
 \label{fig:L_distrib_short}
\end{figure*}

Figure~\ref{fig:L_distrib_short} shows the luminosity distribution of our SGRB population, computed as described in the preceding sections. The thick solid red histogram shows the rate density per logarithmic luminosity bin at redshift zero, $dR_0/d\log_{10}(L_\mathrm{iso}/(\mathrm{erg\,s^{-1}}))$, constructed from our model population, assuming a total local rate $R_0=300\,\mathrm{yr^{-1}\,Gpc^{-3}}$. This local rate has been chosen to make the high luminosity end comparable to estimates by \citealt{Wanderman2015} (W15) and \citealt{Ghirlanda2016} (G16 -- their case \textit{a}). It falls on the low end of the binary neutron star merger rates derived by \citet{TheLIGOScientificCollaboration2018}. The shape of the simulated luminosity distribution is in good agreement with that of W15, which is again remarkable, given that we did not attempt to perform a fitting. In terms of rate, the consistency with W15 would need a lower total rate (around $80\,\mathrm{yr^{-1}\,Gpc^{-3}}$).
The high-luminosity end ($L>10^{52}\mathrm{erg/s}$) is consistent with G16 for this choice of total rate, but the downward turn of the distribution at lower luminosities is not reproduced. \nothing{However, as shown in the next section, a similar feature appears if the jet opening angle at launch is assumed to be narrower.}

\nothing{Figure~\ref{fig:L_distrib_short} reveals another particular feature of our synthetic SGRB population: even the high-end of the luminosity function is dominated by jets seen (slightly) off-axis. Given the uncertainties in the modelling and the degeneration in the parameters (see also next subsection), though, we defer a deeper investigation of this feature to a future work.}

\subsubsection{Dependence on the parameters}

\begin{figure*}[t]
    \centering
    \includegraphics[width=\textwidth]{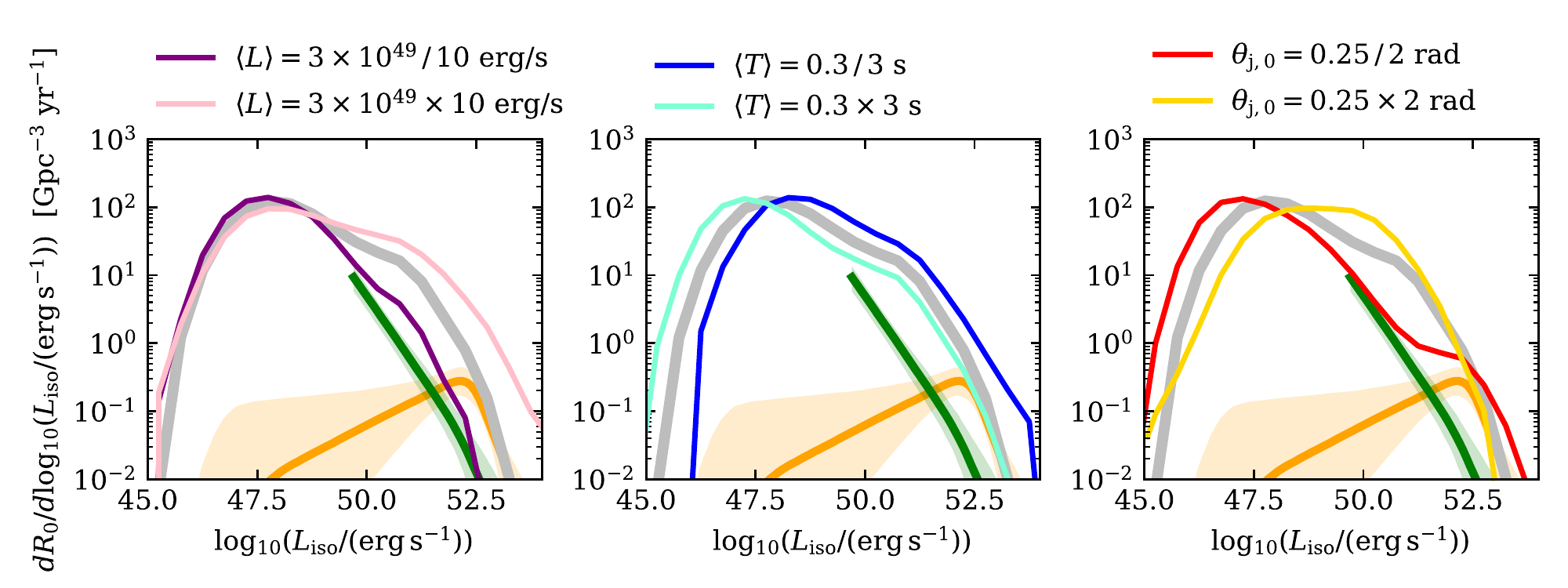}
    \caption{\nothing{Short GRB luminosity function dependence on the parameters of the jet at launch. In each panel, the thick grey line shows the synthetic short GRB luminosity function with fiducial parameters. The thinner, coloured lines show the impact of changing a single parameter. See text for a description of the various dependencies.}}
    \label{fig:short_param_dependence}
\end{figure*}

\begin{figure}
    \centering
    \includegraphics[width=\columnwidth]{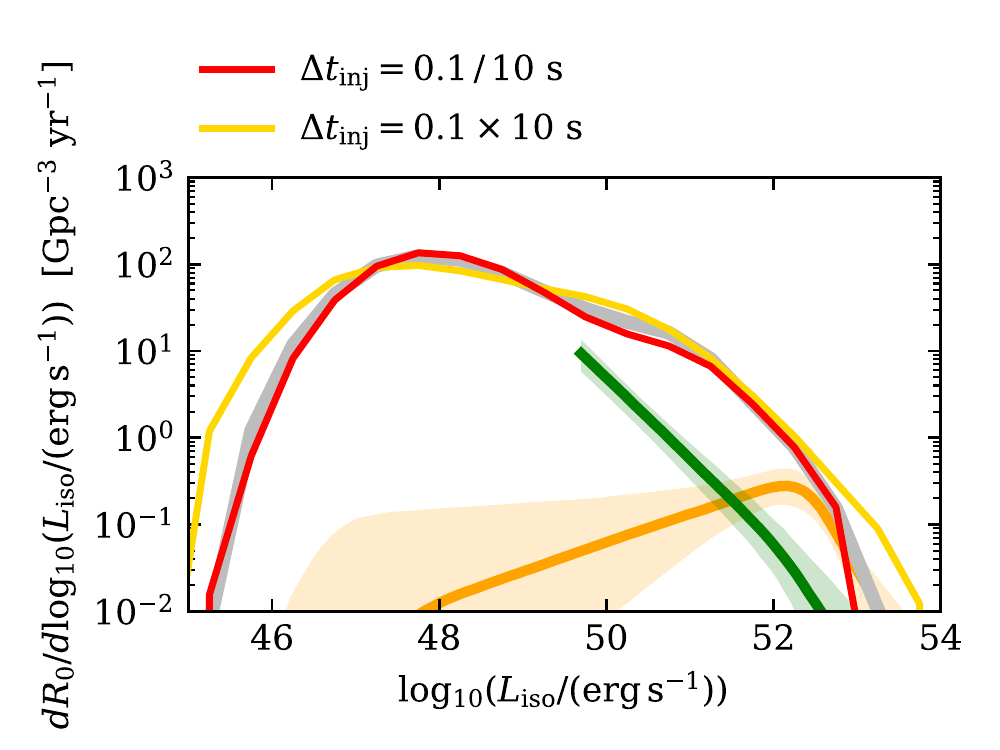}
    \caption{\nothing{Short GRB luminosity function dependence on the delay $\Delta t_\mathrm{inj}$ between merger and jet injection.}}
    \label{fig:short_delta_t_inj_dependence}
\end{figure}

\nothing{Figure~\ref{fig:short_param_dependence} illustrates the dependence of the luminosity function on the jet properties at launch. The meaning of the various curves is the same as in Fig.~\ref{fig:long_param_dependence}. As in the case of long GRBs, most of the trends can be understood intuitively:}
\begin{itemize}
    \item \nothing{similarly to LGRBs, a change in the average jet luminosity $\left < L\right >$ at launch shifts the high-end of the luminosity function correspondingly (see Fig.~\ref{fig:short_param_dependence}, left-hand panel). On the other hand, the low-end, which is dominated by jets seen far off-axis, does not change appreciably. This is due to the homologous expansion of the ambient material \citep{Duffell2018}: jets with a lower luminosity produce slower-moving heads, but as the ambient medium expands these confront themselves with lower densities with respect to heads that move faster. As a consequence, it turns out that the breakout success (and therefore the energy stored in the cocoon after breakout) does not depend explicitly on the jet luminosity (at a fixed initial jet opening angle -- \citealt{Duffell2018});}
    \item \nothing{an increase (resp.~decrease) in the average jet injection duration $\left < T\right >$ shifts the whole luminosity function to lower (resp.~higher) luminosities (central panel of Fig.~\ref{fig:short_param_dependence}). This happens because in this case the jet duration is comparable to the jet breakout time, which has two consequences: (i) the cocoon energy is comparable to that in the jet and (ii) the post-breakout duration is short, which boosts the cocoon luminosity (due to the particular way we model it, see \S\ref{sec:prompt_emission_model}). Since slightly off-axis jets contribute even at the high-end of the luminosity function (with this particular choice of parameters), this impacts the whole luminosity function;}
    \item \nothing{increasing the jet opening angle at launch (yellow line in the right-hand panel of Fig.~\ref{fig:short_param_dependence}) increases the average energy in the cocoon, but it also increases its mass and therefore its opening angle. This shifts the low-luminosity peak of the luminosity function to the right, but leaves the high-end essentially unchanged. On the other hand, a decrease of $\theta_\mathrm{j,0}$ (red line) significantly changes the typical energy structure after breakout, making the jet narrower and more powerful, while the cocoon becomes less energetic. The result is a luminosity function which resembles much more that of LGRBs, where the high-end is dominated by on-axis jets and the low-end by off-axis cocoons. With this choice of parameters, the shape resembles more that of the SGRB luminosity function by \citet{Ghirlanda2016};}
    \item \nothing{Figure~\ref{fig:short_delta_t_inj_dependence} shows the dependence of the luminosity function on the delay $\Delta t_\mathrm{inj}$ between the onset of the expansion of the ambient medium and the start of the jet injection. Reducing the delay by a factor of 10 (red line) does not change the luminosity functionn significantly, since the fiducial value $\Delta t_\mathrm{inj}=0.1\,\mathrm{s}$ is already short compared to the breakout time. Increasing it by the same factor (yellow line) again increases the energy in the cocoon, due to the reduced jet collimation by the ambient medium, which causes the head working surface to widen.} 
\end{itemize}

\section{Discussion}

\subsection{Caveats and open questions}
While the results of our analysis are encouraging about the feasibility of a GRB unification under the quasi-universal jet scenario, we stress that several caveats remain to be addressed. Here is a non-exhaustive list of issues we think need to be investigated in the future:
\begin{itemize}
    \item our model of the jet propagation through the ambient medium, of its collimation by the cocoon, and of the development of structure  builds on previous works \citep{Matzner2003,Bromberg2011,Lazzati2019} which have shown a good agreement with selected numerical simulations. Nevertheless, the model is based on some simplifying assumptions whose impact, to our knowledge, has not been studied in detail yet. To name a few: (i) the jet is assumed to be uniform at launch -- how do the dynamics and the resulting jet structure change if the jet is assumed to be already structured at its base, which is a natural expectation \citep{Kathirgamaraju2019,Christie2019} in a realistic setting? (ii) \nothing{as discussed in \S\ref{sec:jet_propagation},} the inclusion of magnetic field might \nothing{affect} the jet dynamics and collimation \nothing{(but probably not drastically -- \citealt{Levinson2005,Bromberg2016})} -- what is the impact of different degrees of magnetisation and magnetic field configurations on the final jet structure?
    \item There are some indications that the conversion efficiency of jet energy to radiation during the prompt emission might decrease away from the jet axis \citep[e.g.][]{Beniamini2019,Beniamini2019b,Salafia2019} -- how does this impact the luminosity function?
    \item We assumed all progenitors to be identical (separately for LGRBs and SGRBs). How would the population change if we included different progenitors? Would this narrow down even more the required dispersion of jet properties at launch? How do the jet properties at launch depend on the progenitor?
    \item Is it possible to explain the afterglows of LGRBs and SGRBs with a quasi-universal structure seen at different viewing angles?
\end{itemize}
Further theoretical work, and hopefully new precious observations from future detections of off-axis jets, will eventually allow us to address these questions and caveats. 

\subsection{The fraction of choked jets and the duration distribution}

\begin{figure}
    \centering
    \includegraphics[width=\columnwidth]{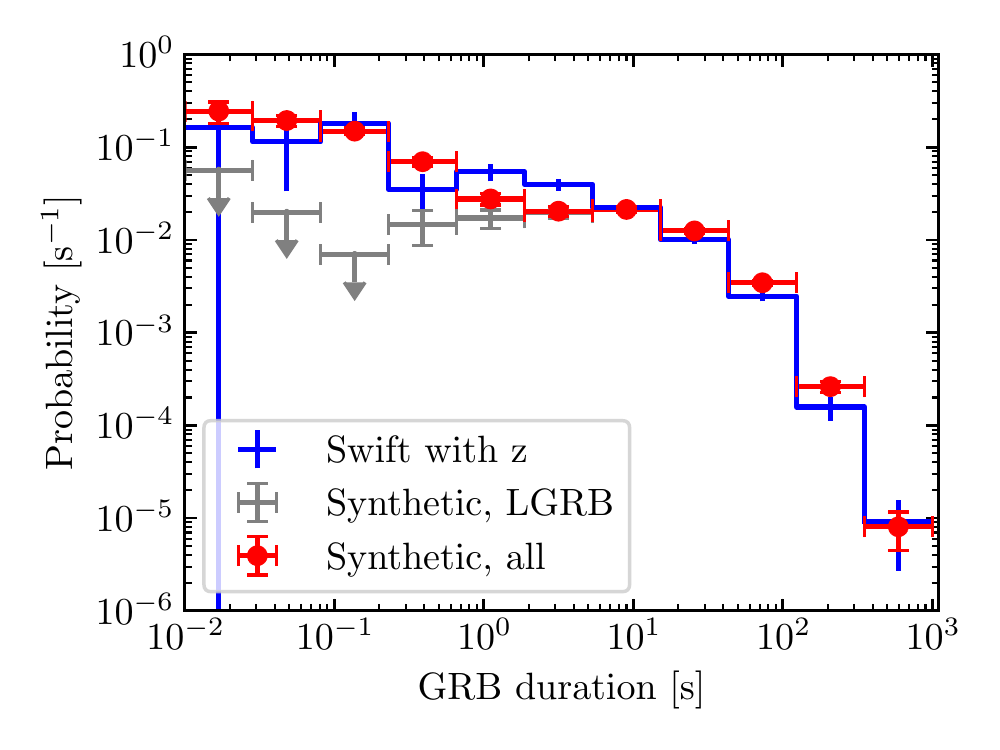}
    \caption{GRB duration probability distribution. The blue histogram shows the estimated probability distribution of intrinsic duration (defined as $T_\mathrm{90}/(1+z)$) of \textit{Swift}-detected GRBs with a measured redshift. Red points represent the corresponding distribution for our simulated populations (assuming an observed rate ratio of 13.5 between LGRB and SGRB, i.e.~the same rate ratio as in the \textit{Swift} sample), while the grey points show the distribution for the LGRB synthetic population alone (downward arrows are upper limits that correspond to empty bins). Vertical error bars show the Poisson uncertainty associated to each bin. }
    \label{fig:duration_distribution}
\end{figure}

During the last few years, some authors \citep{Bromberg2012,Moharana2017,Sobacchi2017,Petropoulou2017} explored the idea that the duration distribution of GRBs could contain information about the minimum time needed for the jet to punch through the typical progenitor. The signature of this should be a flattening in the probability distribution of duration around such minimum time. These works assume an intrinsic power-law probability distribution of jet duration
\begin{equation}
    \frac{dp}{dT}\propto T^{-\alpha}
\end{equation}
above some minimum duration $T_\mathrm{min}$.
Since the jets must spend an average time $T_\mathrm{b}$ to break out of the progenitors, the post-breakout duration ($T_\mathrm{\gamma}$) distribution becomes
\begin{equation}
    \frac{dp}{dT_\mathrm{\gamma}}\propto (T_\mathrm{b}+T_\mathrm{\gamma})^{-\alpha}
\end{equation}
which reflects the intrinsic duration distribution $\propto T_\mathrm{\gamma}^{-\alpha}$ when $T_\mathrm{\gamma}\gg T_\mathrm{b}$, but it is flat when $T_\mathrm{\gamma}\sim T_\mathrm{b}$. The model of \citet{Petropoulou2017} includes a power-law distribution of luminosity, which makes then $T_\mathrm{b}$ luminosity-dependent, but the results are similar. These models reproduce well the observed GRB duration distribution. Since the values of $\alpha$ that reproduce the observed population are large ($\alpha\sim 4$), they predict a large fraction of choked jets (the fraction also depends on the minimum duration $T_\mathrm{min}$ and, in the model of \citealt{Petropoulou2017}, also on the minimum luminosity, but a small fraction of choked jets would require a fine tuning of $T_\mathrm{min}$ close to $T_\mathrm{b}$). In contrast with that, our synthetic populations are dominated by jets that successfully break out from their progenitors (\S\ref{sec:LGRB_injection_properties} and \S\ref{sec:SGRB_injection_properties}). It is natural, therefore, to ask whether the duration distribution of our populations does reflect the actual observed one. Figure~\ref{fig:duration_distribution} shows a comparison between the observed duration probability distribution and that of our synthetic sample. The blue histogram shows the estimated probability distribution of rest-frame duration (defined as $T_\mathrm{90}/(1+z)$) of the sample of \textit{Swift} GRBs with measured redshift. For each bin, we compute the probability estimate as
\begin{equation}
    \frac{dp}{dT}(T_i)\approx \frac{N_i}{N_\mathrm{tot}\,\Delta T_i},
\end{equation}
where $T_i$ is the central duration of the bin, $\Delta T_i$ is the bin width, $N_i$ is the number of events in that bin, and $N_\mathrm{tot}$ is the total number of events. The associated (Poisson) uncertainty is
\begin{equation}
    \delta \frac{dp}{dT}(T_i)\approx \frac{\sqrt{N_i}}{N_\mathrm{tot}\,\Delta T_i}.
\end{equation}
For empty bins, we take as upper limit the estimate obtained by assuming a single event in that bin.
Red points show the corresponding probability distribution (for the duration defined in \S\ref{sec:prompt_emission_model}) of our synthetic samples, where we assumed a 13.5 observed rate ratio of LGRB over SGRB (i.e.~the same as for the \textit{Swift} sample used). Grey points show the LGRB synthetic sample alone. The comparison shows a good agreement, indicating that the observed GRB duration distribution is still consistent with the possibility that most GRB jets successfully break out of their progenitor. The observation of a successful GRB jet associated to GW170817 \citep{Ghirlanda2019,Mooley2018} suggests that this is the case for SGRBs \citep[see][for an excellent discussion of this latter point]{Beniamini2019}.

\section{Conclusions}

During the last two decades, evidence accumulated in favour of the presence of jet structure in both LGRB and SGRB, unveiling its important role in shaping the appearance of these sources. The limited range of properties of their putative progenitors seems to suggest the possibility that such jet structure is quasi-universal: this would have a great unifying impact. In this work, we set up a model of the interaction of a GRB jet with its environment, specifically aimed at predicting the jet structure after breakout. With this model, we computed self-consistently the quasi-universal structure that would result if the properties of GRB jets at launch and those of their progenitors were distributed in very narrow ranges. Strikingly, these quasi-universal structures can reproduce both the LGRB and SGRB luminosity functions. While we acknowledge that several issues remain to be addressed, we consider this results is encouraging towards the unification of GRBs within the quasi-universal hypothesis.

\begin{acknowledgements}
\nothing{We thank the anonymous referee for raising important points that helped to improve the quality of this work.}  We thank R.~Ciolfi for the discussion that sparked the research behind this work. We thank G.~Ghirlanda and G.~Ghisellini for stimulating discussions and for their continued support and motivation. We thank D.~Lazzati and B.~Giacomazzo for insightful comments on the first draft. O.~S.~acknowledges the MIUR (Italian Ministry of University and Research) grant ``FIGARO'' (1.05.06.13) for support. S.~A.~acknowledges the GRAvitational Wave Inaf TeAm - GRAWITA (P.I. E. Brocato) for support.
\end{acknowledgements}

\footnotesize{
\bibliographystyle{aa}
\bibliography{references}
}

\appendix

\section{Comparison with simulations}
\label{sec:comparison_simulations}
In this section we compare our semi-analytical model with results of a number of numerical relativistic hydrodynamical simulations described in the literature.

\subsection{Jet head propagation}\label{sec:comparison_simulations_head}

\begin{table}
    \centering
    \begin{tabular}{ccccc}
        \citealt{Gottlieb2018a} & \multicolumn{2}{c}{$t_\mathrm{bo}$ [s]} & \multicolumn{2}{c}{$E_\mathrm{c}$ [$10^{49}$ erg]}   \\
        \midrule
        Configuration & Sim. & Model & Sim. & Model \\
        A & 0.10 & 0.10 & 1.5 & 1.4 \\
        B & 0.20 & 0.18 & 2.0 & 2.2 \\
        C & 0.12 & 0.14 & 1.5 & 1.7 \\
        \midrule
        \citealt{Nagakura2014} & \multicolumn{2}{c}{$t_\mathrm{bo}$ [s]} & \multicolumn{2}{c}{$r_\mathrm{bo}$ [$10^{9}$ cm]}   \\
        \midrule
        Configuration      &   Sim. & Model & Sim.  & Model \\
        M-ref   &	0.23 & 0.24  & 	3.7  &  3.5  \\
        M-L4    &	0.20 & 0.18  & 	3.2  &  2.7  \\
        M-th30  &	0.63 & 0.55  & 	8.9  &  7.2  \\
        M-th45  &	--   & 1.01  & 	--   &  13   \\
        M-ti500 &	0.90 & 0.69  &  17.5 &  14   \\
        M-M3    &	0.11 & 0.10  & 	2.0  &  1.7  \\
        M-M2-2  &	0.32 & 0.32  & 	5.0  &  4.4  \\
        M-M1    &	0.75 & 0.64  & 	11.0 &  8.2  \\  
         
    \end{tabular}
    \caption{Comparison between the breakout time, cocoon energy and breakout radius predictions of our model and those found in numerical simulations of \cite{Gottlieb2018a} and \citet{Nagakura2014}. The energies reported in \cite{Gottlieb2018a} refer to both the jet and the counter-jet cocoons, so they have been divided by 2.}
    \label{tab:gottlieb_comparison}
\end{table}

Our modelling of the jet head propagation follows quite closely previous works \citep{Matzner2003,Bromberg2011}, with some adjustment to account for a moving ambient medium. Similar models have been presented recently by \citet{Matsumoto2018}, \citet{Gill2019} and \citet{Lazzati2019}, with some minor differences. We investigate here the validity of this approach by comparing the breakout time, breakout radius and cocoon energy predicted by our model to those obtained by \citet{Gottlieb2018a} (G18 hereafter) and \citet{Nagakura2014} (N14 hereafter) in numerical simulations. Table~\ref{tab:gottlieb_comparison} shows the results of our comparison. 

Configurations A, B and C in G18 all involve a relativistic jet of luminosity $L_\mathrm{j}=2\times 10^{50}\,\mathrm{erg/s}$ and initial opening angle $\theta_\mathrm{j,0}=10^\circ$ launched in an ambient medium representing binary neutron star ejecta in homologous expansion, with a total mass $M_\mathrm{ej}=10^{-2}\,M_\odot$ and a power-law density profile decreasing as $r^{-3.5}$. The configurations differ by the maximum velocity $\beta_\mathrm{max}$ of the ejecta (being $0.4$ in configuration C and $0.2$ in the other two) and by the delay $t_\mathrm{inj}$ between the start of the ejecta expansion and the jet injection ($t_\mathrm{inj}=80$, $240$ and $40$ ms for A, B and C respectively). We reproduce the ambient medium configuration and follow the jet evolution semi-analytically using our model, up to the jet breakout. The breakout times and cocoon energies predicted by our model agree with the simulations within 20\%. 

The initial configurations of N14 are very similar to those of G18. Their reference configuration (M-ref) is the same as case C of G18, but with $t_\mathrm{inj}=50\,\mathrm{ms}$ and $\theta_\mathrm{j,0}=15^\circ$. The other configurations, listed in Table~\ref{tab:gottlieb_comparison}, are variants over the reference one, exploring alternative values of some parameters, namely $L_\mathrm{j}=4\times 10^{50}\,\mathrm{erg/s}$ (M-L4), $\theta_\mathrm{j,0}=30^\circ$ (M-th30) and $45^\circ$ (M-th45), $t_\mathrm{inj}=500\,\mathrm{ms}$ (M-ti500), and $M_\mathrm{ej}=10^{-3}\,M_\odot$ (M-M3), $2\times 10^{-2}\,M_\odot$ (M-M2-2) and $10^{-1}\,M_\odot$ (M-M1). Also in this case, the results of our model agree with the simulations within $\sim 20\%$, but they show a systematic trend towards faster breakouts, which may indicate that the $\tilde L$ correction adopted from \citet{Harrison2018} needs to be modified in the case of a moving ambient medium.

\subsection{Collimation and jet opening angle} \label{sec:comparison_simulations_collimation}

\begin{figure*}
    \centering
    \includegraphics[width=\textwidth]{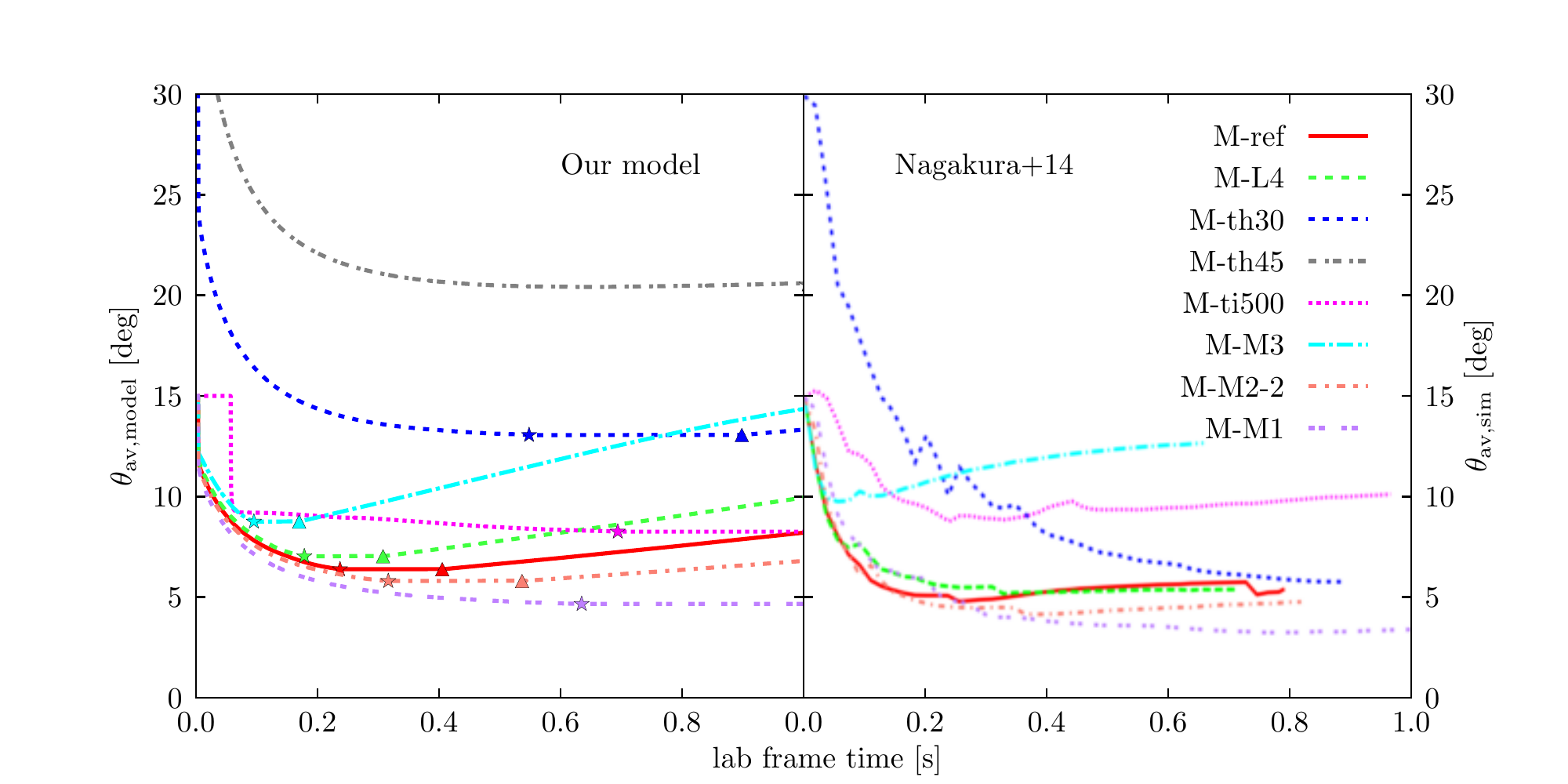}
    \caption{Comparison between the height-averaged opening angle evolution from the numerical simulations in \citealt{Nagakura2014} (right-hand panel, adapted from their Figure~3) and those predicted by our semi-analytical model (left-hand panel). Each line refers to a different initial configuration (see text). Stars and upward triangles in the left-hand panel mark $t_\mathrm{bo}$ and $t_\mathrm{bo}+t_\mathrm{delay}$ (see \S\ref{sec:Jet_structure}), respectively. The evolution for configuration M-th45 is not shown in the original figure.}
    \label{fig:nagakura_angles}
\end{figure*}

We perform an additional comparison with the time evolution of the average jet opening angles as reported in N14. Figure~\ref{fig:nagakura_angles} shows the time evolution of the height-averaged opening angle of the jet from N14 (bottom panel) and as predicted by our model (top panel). The angle is computed as
\begin{equation}
    \theta_\mathrm{av}(t) = \frac{\int_{z_0}^{z_\mathrm{h}(t)}\theta_\mathrm{j}(z,t)dz}{z_\mathrm{h}(t)-z_0}
\end{equation}
In the simulations, $\theta_\mathrm{j}(z,t)$ is the opening angle containing the ``relativistic'' ($h\Gamma>10$) material located at a height $z$, at time $t$. In our model, according to our assumptions detailed above, the jet is conical up to $Z=(\hat{z}+z_c)/2$ and cylindrical above that point, so that in our case
\begin{equation}
    \theta_\mathrm{av}(t)=\frac{\theta_\mathrm{j,0}Z(t)}{z_\mathrm{h}(t)-z_0}\left[1-\frac{z_0}{Z(t)} + \ln\left(\frac{z_\mathrm{h}(t)}{Z(t)}\right)\right]
\end{equation}
before breakout ($t\leq t_\mathrm{bo}$), and
\begin{equation}
    \theta_\mathrm{av}(t)=\frac{\theta_\mathrm{j}(t)z_\mathrm{bo}}{z_\mathrm{bo}-z_0}\left[1-\left(\frac{\theta_\mathrm{j,0}}{\theta_\mathrm{j}(t)}\right)\frac{z_0}{z_\mathrm{bo}} + \ln\left(\frac{\theta_\mathrm{j,0}}{\theta_\mathrm{j}(t)}\right)\right]
\end{equation}
after breakout ($t > t_\mathrm{bo}$).
Figure~\ref{fig:nagakura_angles} shows a remarkably good agreement (given the simplifications adopted) between our model and the simulations in most cases, after setting $\alpha=1/8$ in Eq.~\ref{eq:theta_j(t)}. With the initial configuration M-th45 (larger jet)  \cite{Nagakura2014} find that their jet fails to penetrate the ejecta within their simulation time $1\,\mathrm{s}$, so they do not report the breakout time and radius. The imperfect agreement in case M-th30 suggests that the small angle approximations adopted in the treatment of collimation may break down at jet opening angles as large as $\gtrsim 30^\circ$.

\subsection{Jet and cocoon structure}\label{sec:comparison_simulations_structure}

In order to compare our recipe for the jet structure to numerical simulations, we reproduce the charachteristics of the ambient media and the properties of the jet at its base from the simulations described in \citet{Lazzati2017} and \citet{Xie2018}. We then apply our semi-analytical model of the jet head propagation (\S\ref{sec:jet_propagation}) to find the breakout quantities. Finally, we use the latter as input to compute the jet and cocoon structures as detailed in \S\ref{sec:Jet_structure}, \S\ref{sec:Cocoon_structure} and \S\ref{sec:Gamma_structure}. Let us describe the three configurations and the results of the comparison.

\subsection*{Lazzati et al. 2017}

\begin{figure}
 \includegraphics[width=\columnwidth]{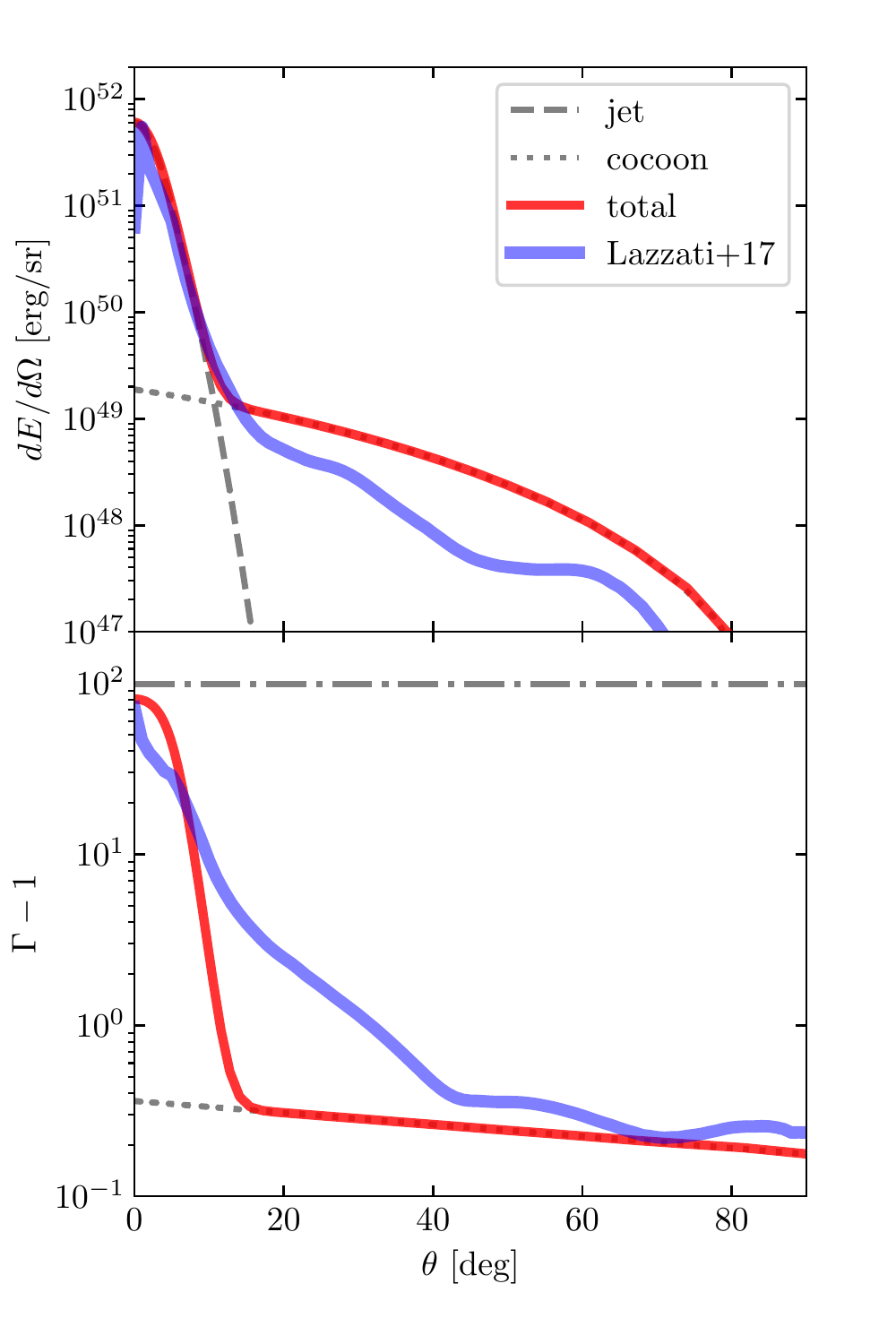}
 \caption{Comparison between the jet structure predicted by our model and the results of numerical simulations by \citet{Lazzati2017}. The upper panel refers to the distribution of kinetic energy per unit solid angle, while the lower panel shows the average Lorentz factor diminished by one. Grey dashed and dotted lines represent respectively the jet and the cocoon as computed by our model (\S\ref{sec:Jet_structure}, \S\ref{sec:Cocoon_structure} and \S\ref{sec:Gamma_structure}). Red lines show the combined structure. Blue thick lines show the results of the numerical simulations (the Lorentz factor profile is given in \citealt{Lazzati2018}).}
 \label{fig:Lazzati_comparison}
\end{figure}

\citet{Lazzati2017} used a modified version of the relativistic hydrodynamics code \texttt{FLASH} \citep{Fryxell2000} to simulate a jet propagating and breaking out of an ambient medium with a static, spherically symmetric density distribution with a profile $\rho(r)\propto r^2\exp(-r)$. The jet is injected within an angle $\theta_\mathrm{j,0}=16^\circ$ with a constant luminosity $L_\mathrm{j}=10^{50}\mathrm{erg/s}$ and lasts for $T_\mathrm{jet}=1\,\mathrm{s}$. 
Even though the authors state that the terminal Lorentz factor of the injected jet is $\Gamma_\mathrm{300}$, the actual maximum Lorentz factor attained in the simulation is lower, presumably because its evolution was not followed up to the coasting phase. We thus choose to set $\Gamma_\mathrm{j}=100$. The resulting jet structure is shown in Figure~\ref{fig:Lazzati_comparison}. The general agreement is good, especially for what concerns the kinetic energy density distribution. The Lorentz factor profile is in quite good agreement in the inner jet and in the outer cocoon, but it falls off too fast in the transition from the jet to the cocoon.

\subsection*{Xie \& MacFadyen 2018}

\begin{figure}
 \includegraphics[width=\columnwidth]{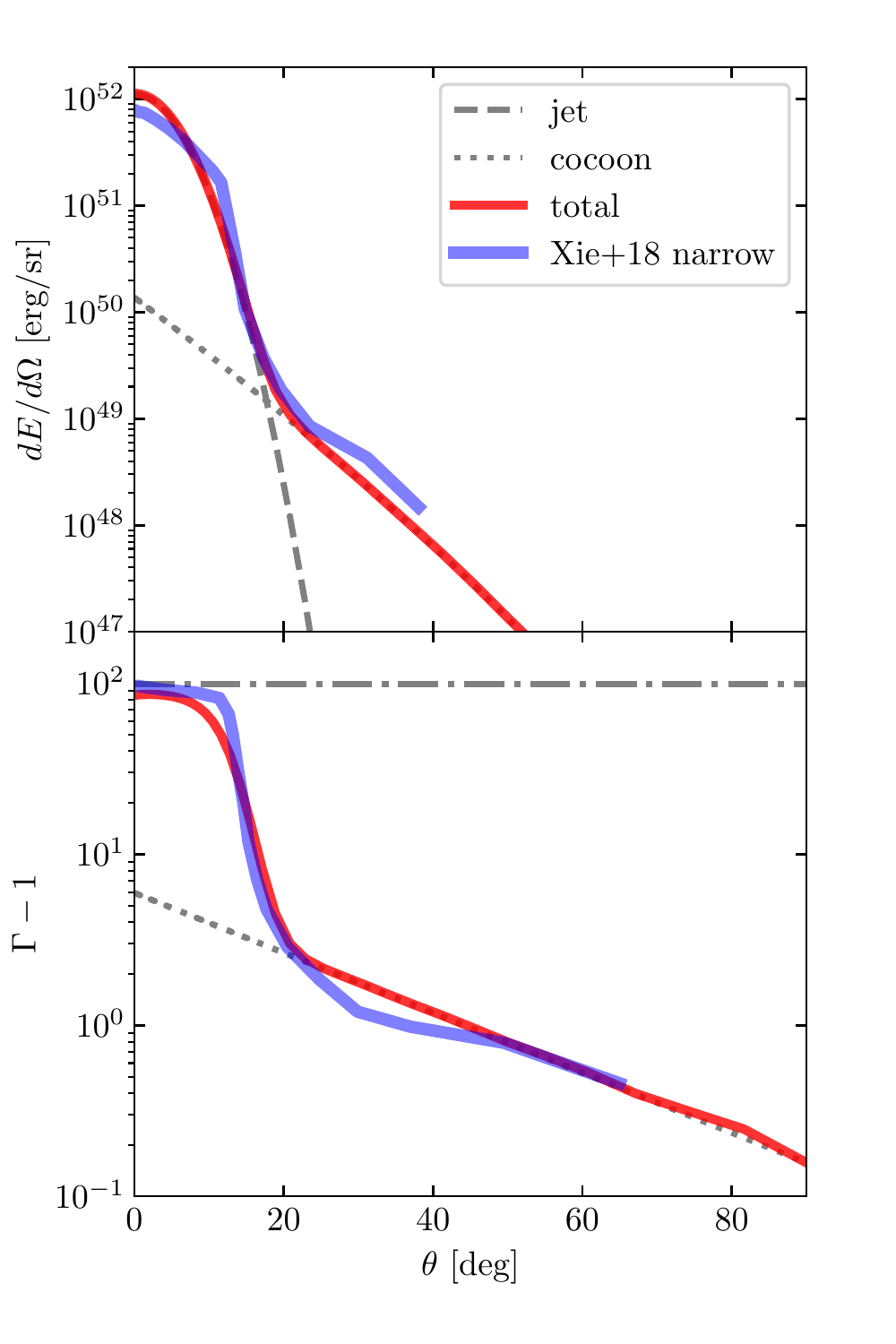}
 \caption{Same as Fig.~\ref{fig:Lazzati_comparison}, but for the ``narrow engine'' case in \citet{Xie2018} (we choose the structure extracted from the $10^4\,\mathrm{s}$ snapshot as shown in their Figure~2, right-hand panels).}
 \label{fig:Xie_comparison_narrow}
\end{figure}
\begin{figure}
 \includegraphics[width=\columnwidth]{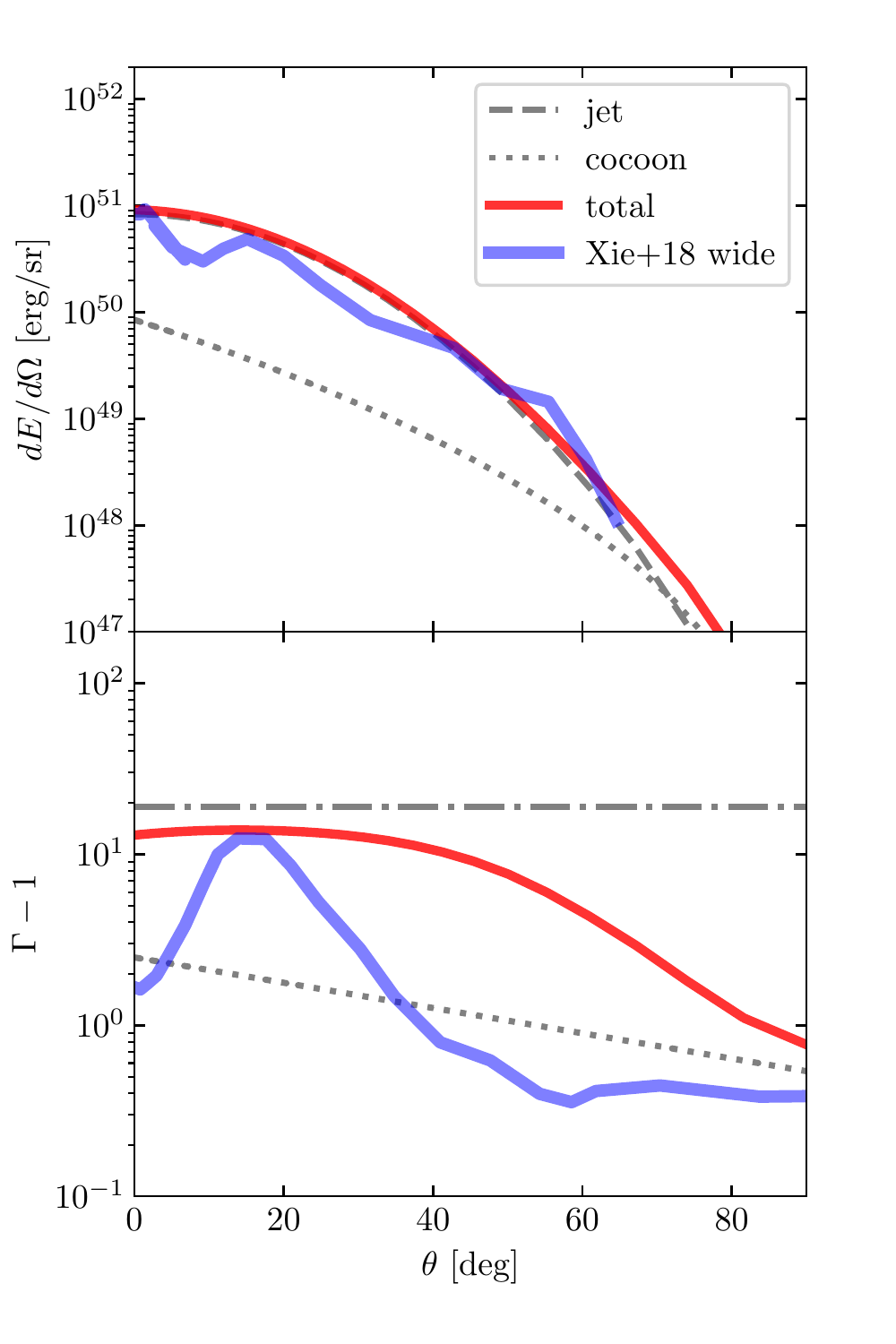}
 \caption{Same as Fig.~\ref{fig:Lazzati_comparison}, but for the ``wide engine'' case in \citet{Xie2018} (we choose the structure extracted from the $10^4\,\mathrm{s}$ snapshot as shown in their Figure~7, right-hand panels). }
 \label{fig:Xie_comparison_wide}
\end{figure}

The simulations in \citet{Xie2018} model two jets propagating in a moving ambient medium which represents the result of the merger of two neutron stars. The medium is in homologous expansion and features an inner denser cloud surrounded by a fast tail with a steep density fall-off. The two cases modeled, dubbed ``narrow engine'' and ''wide engine'', differ by the opening angle of the jet at its base, and by the jet enthalpy at injection. One subtlety is that the jets in the two models are injected following different techniques: in the ``wide engine'' case, the jet is injected through a cylindrical nozzle and thus has a quite well-defined opening angle $\theta_\mathrm{j,0}=0.35\,\mathrm{rad}$; in the ``narrow engine'' case, the jet is injected through a circular nozzle as in \citet{Duffell2015}, and features a non-uniform angular distribution of luminosity, with an approximately Gaussian form $\simpropto \exp[-(\theta/\theta_\mathrm{0})^2/2]$, with $\theta_0=0.1\,\mathrm{rad}$. We thus try different values of the base jet half-opening angle $\theta_\mathrm{j,0}$ in our model between $\theta_0$ and $2\theta_0$, looking for the best agreement with their results. We find \nothing{that} $\theta_\mathrm{j,0}=1.38\theta_0$ produces a very good agreement in both the kinetic energy and Lorentz factor profile (setting $\Gamma_\mathrm{j}=100$, i.e.~equal to their terminal Lorentz factor), as shown in Fig.~\ref{fig:Xie_comparison_narrow}. In the ``wide engine'' case this tuning is not necessary, and we simply adopt the value $\theta_\mathrm{j,0}=0.35\,\mathrm{rad}$ as in the simulation, and we set $\Gamma_\mathrm{j}=20$, i.e.~again equal to their terminal Lorentz factor. The result is shown in Fig.~\ref{fig:Xie_comparison_wide}. In this latter case, the  agreement in the Lorentz factor profile is not as good.

These comparisons show that, while somewhat crude, our recipe seems to capture the main trends in the development of the jet and cocoon structure, at least in the conditions covered by these simulations. We plan to run a series of dedicated numerical hydrodynamical simulations to investigate further these aspects.

\end{document}